\documentclass[graybox, nosecnum]{svmult}

\usepackage{xspace}
\usepackage{mathptmx}       
\usepackage{helvet}         
\usepackage{courier}        
\usepackage{type1cm}        
\usepackage{amsmath,amssymb}
\usepackage{makeidx}         
\usepackage{graphicx}        
\usepackage{multicol}        
\usepackage[bottom]{footmisc}
\usepackage{caption}
\usepackage{subcaption}
\usepackage{hyperref}        
\usepackage{soul}            
\hypersetup{colorlinks=true,urlcolor=blue}
\usepackage[]{natbib}

\newcommand{\nustar}{\textit{NuSTAR}\xspace}
\newcommand{\ixpe}{\textit{IXPE}\xspace}
\newcommand{\xte}{\textit{RXTE}\xspace}
\newcommand{\nicer}{\textit{NICER}\xspace}

\newcommand{\psig}{\ensuremath{P_{\rm s}}\xspace}
\newcommand{\pmeas}{\ensuremath{P}\xspace}
\newcommand{\deadt}{\ensuremath{\tau_{\rm d}}\xspace}
\newcommand{\zsq}{\ensuremath{Z^2_n}\xspace}
\newcommand{\zsqbin}{\ensuremath{Z^2_{n,\mathrm{bin}}}\xspace}

\newcommand{\nbin}{\ensuremath{N_{\rm bin}}\xspace}

\newcommand{\chizsq}{\ensuremath{\chi^2_{2n}}\xspace}

\newcommand{\eref}{Equation~\ref}

\makeindex             

\begin{document}
\title*{Fourier Domain}
\author{Matteo Bachetti \thanks{corresponding author} and Daniela Huppenkothen}
\institute{Matteo Bachetti \at INAF-Osservatorio Astronomico di Cagliari, via della Scienza 5, 09047 Selargius (CA), Italy, \email{matteo.bachetti@inaf.it}
\and Daniela Huppenkothen \at SRON Netherlands Institute for Space Research, Niels Bohrlaan 4, 2333CA Leiden, The Netherlands \email{d.huppenkothen@sron.nl}}
\maketitle
\abstract{The changes in brightness of an astronomical source as a function of time are key probes into that source's physics. Periodic and quasi-periodic signals are indicators of fundamental time (and length) scales in the system, while stochastic processes help uncover the nature of turbulent accretion processes. A key method of studying time variability is through Fourier methods, the decomposition of the signal into sine waves, which yields a representation of the data in frequency space. With the extension into \textit{spectral timing} the methods built on the Fourier transform can not only help us characterize (quasi-)periodicities and stochastic processes, but also uncover the complex relationships between time, photon energy and flux in order to help build better models of accretion processes and other high-energy dynamical physics. In this Chapter, we provide a broad, but practical overview of the most important relevant methods.}

\section{Keywords}
Fourier analysis; Time series analysis; X-ray oscillations; Pulsars; Spectral Timing; Time lags;
\tableofcontents{}

\section{Introduction}

Astrophysical compact objects show variable signals of all kinds and on a very broad range of timescales, ranging from milliseconds to decades.
Some modulations are very coherent and stable, such as the pulsation produced by the rotation of some millisecond pulsars or the signatures of orbital periods.
Others are quasi-periodic, meaning that these signals vary their frequency or phase unpredictably in a relatively short time scale around a specific frequency: this is the case of quasi-periodic oscillations in black hole (BH) and neutron star (NS) binaries \citep{vanderklisRapidXrayVariability2006}, or quasi-periodic eruptions in AGN \citep{miniutti2019}.
Many sources also show some stochastic variability, some on long timescales of years to decades.
While for many sources, their behaviour changes on timescales ranging from weeks to decades, there exists a class of phenomena called \textit{fast transients} that brighten and dim on comparatively short timescales of milliseconds to minutes, such as fast radio bursts (FRB; \citealt{petroffFastRadioBursts2022}), X-ray bursts \citep{Galloway2021}, magnetar bursts \citep{reaMagnetarOutburstsObservational2011}.

While all these phenomena require dedicated techniques of investigation and one-size-fits-all solutions are rare, there is a broad class of methods that have been adapted to and successfully applied in a broad range of contexts: Fourier analysis. These methods are based on the Fourier transform, a decomposition of a function in the time domain into a representation as a sum of sines and cosines in the frequency domain (a short formal definition will follow in the next Section, and a more in-depth discussion of the underpinning mathematical background can be found in \citealt{belloniBasicsFourierAnalysis2022}). The \textit{periodogram}, a measure of how much variability or ``power'' exists in the data at a given temporal frequency, is particularly helpful in  characterizing stochastic processes, which we will describe in more detail along with other typical types of variability further along in this Chapter, as well as in finding (quasi)periodic signals in the data.

Over the past decades, an understanding has emerged that for many accreting sources, temporal and energy variations are intricately linked. For example, high-energy photons produced in the accretion flow close to a stellar-mass black hole are reprocessed and reflected in the optically thick, geometrically thin accretion disk, a process observed as a characteristic reflection signature in X-ray spectra. This reflection can also be studied through \textit{time lags}: for example, the light emitted at high energies (as a tracer of the corona near the black hole) exhibits stochastic changes in flux, and these temporal patterns can also be observed at lower photon energies (a tracer of the accretion disk), but shifted in time to arrive at the telescope slightly after the high-energy emission. This is a signature of the additional light travel time of the reprocessed photons as they impinge on the disks, are reprocessed, and then re-emitted. Similarly, one can trace the path of turbulent density patterns from the outer to the inner regions of the accretion disk through time lags as well, but in the other direction (the lower energy emission arrives first). These mechanisms, along with others, can be probed through \textit{spectral timing}, a class of statistical methods that extend the traditional Fourier transform to take the two-dimensional time-energy data provided by most major X-ray telescopes and compute informative representations of that data that can probe specific behaviours. Examples include the cross spectrum, the related concepts of coherence and time lags, and time-energy representation like variability spectra and will be the subject of the second half of this Chapter.
We will end this chapter with a discussion of common pitfalls in analysing X-ray data with Fourier methods, most notably detector effects like dead time and irregular sampling, and provide a brief outlook into the future.

As a final note, it is worth mentioning that different branches of science and engineering use different terminology, normalizations and conventions when calculating and working with periodograms.
Here, we will describe the typical use in X-ray astronomy, giving pointers, when relevant, to equivalent methods and terms used in other branches of science. At the same time, while this chapter is geared towards providing a comprehensive introduction into the method relevant to studying astronomical sources in X-rays, Fourier methods have broad applicability both within and beyond astronomy, and we hope this chapter provides a useful introduction beyond its intended audience.

\section{Fourier Basics}
\label{sec:fourierbasics}
\subsection{Terminology and Notation}

It is helpful to introduce a number of terms that will be relevant for the rest of this chapter. In most of what we consider in this chapter, the starting point will be a data set comprising measurements of brightness (such as flux, or in X-rays, often photon counting data) $x$ as a function of time\footnote{It is worth keeping in mind that much of what we describe in the chapter is also applicable when the dependent variable is not time, but some other quantity, but for our purposes, we will stick with time as the dependent variable.} $t$. Such a data set is called a \textit{time series}, though in astronomy, the term \textit{light curve} is more common. Flux measurements in astronomy are almost never instantaneous; instead, one might integrate over a time $\Delta t$ to generate a flux measurement. X-ray instruments generally produce \textit{event lists}, i.e.~data sets comprising of individual recorded events, often with a time of arrival and a photon energy (and potentially additional information) attached. Where not stated otherwise, it is assumed that these event lists have been \textit{binned} into time series, recording the number of events with arrival times in each time bin as well as either the start time or the mid-point of each time bin. In this case, the width of each time bin $\Delta t$ can be chosen by the researcher, subject to the constraint that it must be wider than the precision of the instrument in measuring photon arrival times.

As a result, the starting point for most of our explorations is a list of $N$ counts or fluxes $\{x_i\}_{i=1}^{N}$ and an associated list of time bin mid-points $\{t_i\}_{i=1}^{N}$, which form a time series of total duration $T$ and with a time resolution of $\Delta t = T/N$. For much of this Chapter and unless otherwise mentioned, $\Delta t$ is assumed to be constant and equal for all time bins. For spectral timing, which we will consider later in the chapter, the event list might first be split into energy bins\footnote{Photon energy of individual photon events is often recorded in discrete bins called \textit{channels}, which may be further binned to generate physically interesting energy bands}, such that time series $\{t_{i,k}, x_{i,k}\}_{i=1}^{N}$ contains all events recorded in energy bin $k$ between $E_{k, \mathrm{min}}$ and $E_{k, \mathrm{max}} = E_{k, \mathrm{min}} + \Delta E$.

A note on terminology in the context of this work: while often used interchangeably, in this work we will use the term \textit{periodogram} (also sometimes known as the \textit{Schuster periodogram}, \citealt{schuster1898}) to denote a set of powers derived from an observed (or simulated) time series, that is, a realization of a (stochastic) process. We will use \textit{power spectrum} or \textit{power spectral density (PSD)} to denote the underlying process that generated this data. The difference between the two will become more obvious in a later section on stochastic processes.

\subsection{The Periodogram}

This section will serve as a brief primer on the Fourier transform and periodograms, and will set up relevant notation for the remainder of the chapter but not go into particular depth (see, e.g.~\cite{vanderklisFourierTechniquesXray1989} for a more in-depth treatment). As we have set up above, let us assume we measure a set of $N$ observations $\{x_i\}_{i=1}^N$ at equidistantly spaced points in time $\{t_i\}_{i=1}^{N}$.

The (discrete) Fourier transform decomposes the series of observations $\{x_i\}_{i=1}^{N}$ into sines and cosines. Intuitively, one represents the observed time series with a sum of $K$ sines and cosines across a range of frequencies, each with an amplitude $A_k$ and a phase $\phi_k$. More formally, the Fourier transform of the series $\{x_i\}_{i=1}^{N}$ can be written as

\begin{equation}\label{eq:ft}
    a_k = \sum_{i}x_i \exp{(2\pi i \nu_k t_i)}\;;
\end{equation}
the Fourier amplitudes $a_k$ are complex numbers describing the amplitude and phase of the sine wave representation at frequency $\nu_k$.

Equivalently, we can express the series of observations through the inverse transform:
\begin{equation}
    x_i = \frac{1}{N} \sum_{k}a_k \exp{(-2\pi i \nu_k t_i)}
\end{equation}

Please note that many widely used computational libraries mix up the definitions, either inverting the signs of the exponentials or assigning the factor $1/N$ to the direct transform instead of the inverse transform.
One such example is the \texttt{numpy} Python library \citep{harris2020array}, whose \texttt{numpy.fft} module inverts the sign of the two transforms.

The set of frequencies $\nu_k$ can, in principle, be freely chosen, but in practice it is generally advisable to choose a set of frequencies that adequately represents the information contained in the data. This suggests that the range of frequencies that should be evaluated is set by the total length $T$ of the time series, as well as the spacing between data points $\Delta t$. The former sets the smallest frequency that can be probed and correspondingly frequency resolution $\Delta \nu$: signals with periods longer than the total duration of the time series cannot be resolved. Conversely, signals with periods shorter than $2\Delta t$ are cannot be resolved either, and thus defines the \textit{Nyquist frequency} as $\nu_{\mathrm{Ny}} = 0.5 / {\Delta t}$. For a real-valued function, the discrete Fourier transform is symmetric about the origin. That, together with the constraints on the frequency resolution and the Nyquist frequency, suggests that there are $N/2$ (uncorrelated) frequencies. \textit{Oversampling}, i.e.~calculating the Fourier transform for a denser grid of frequencies, while possible, is not recommended for the applications considered in this chapter, because they complicate the statistical treatment of the resulting periodogram by introducing correlations between neighbouring powers.

The periodogram is defined as the squared modulus of the Fourier transform:
\begin{equation} \label{eq:periodogram}
    P_k = {|a_k|}^2.
\end{equation}
Because the phase shift between a time series and itself is zero, the result is a set of real amplitudes measuring the \textit{power} $P_k$ at frequency $\nu_k$, a measure of the amplitude of the sine wave at that frequency required to represent the data. Parseval's theorem implies that the sum of powers corresponds to the total variance in the time series,

\begin{equation}\label{eq:var}
    \mathrm{Var}(x_i) = \frac{1}{N}\sum_{k=-N/2, k \neq 0 }^{N/2}|a_k|^2 \; .
\end{equation}

\noindent Note that the zeroth frequency is not part of this calculation, because it encodes the integral of the flux in the time series (in X-rays, corresponding to the total number of photons).

There are a number of useful ways to normalize the periodogram for specific applications. The \textit{Leahy normalization} \citep{leahySearchesPulsedEmission1983} is one commonly used normalization that is particularly useful when comparing signals against an instrumental noise floor (such as when searching for periodic signals, as we will see later) because said instrumental noise will be normalized to always have the same mean power and statistical properties. It is defined as

\begin{equation}\label{eq:leahy}
    P_{k, \mathrm{Leahy}} = \frac{2}{N_\mathrm{ph}} |a_k|^2, \; \; k= 0, \dots, \frac{N}{2}
\end{equation}

\noindent where $N_\mathrm{ph}$ corresponds to the total number of photons for photon counting data.
Note that for Poisson white noise, $N_\mathrm{ph}=N\mu_\mathrm{ph}=N \sigma{x}^2$, where $\mu$ is the mean flux of the light curve (in counts/bin if it is obtained by counting events) and $\sigma{x}^2$ is the average variance of light curve measurements.
So, in principle, an equivalent normalization can be obtained for non-Poisson data, by substituting $N_\mathrm{ph}$ with $N \mathrm{Var}(x_k)$:
\begin{equation}\label{eq:leahygauss}
    P_{k, \mathrm{Leahy, gauss}} = \frac{2}{N \sigma{x}^2} |a_k|^2, \; \; k= 0, \dots, \frac{N}{2}
\end{equation}

Another normalization, the root-mean-square or r.m.s. normalization \citep{belloniAtlasAperiodicVariability1990,miyamotoCanonicalTimeVariations1992} is particularly valuable when one is interested in the variability, normalized by the mean flux, in a part of the periodogram (like for example, in a QPO, as we will see further along).
In this normalization the integrated power over a given frequency range describes the fractional variability in that range:

\begin{equation}\label{eq:rms}
    P_{k,\mathrm{rms}} = \frac{2 T}{(N \mu_\mathrm{ph})^2} |a_k|^2 \; ,
\end{equation}

\noindent where $T$ is the total length of the time series (and $\mu_\mathrm{ph}$, as above, the count rate in counts/bin).
This can be corrected by the background level, yielding the \textit{source fractional rms}:
\begin{equation}\label{eq:srcrms}
    P_{k,\mathrm{source\,rms}} = \frac{2 T}{N^2( \mu_\mathrm{ph} - \mu_\mathrm{back})^2} |a_k|^2 \; ,
\end{equation}

Another oft-used normalization measures the \textit{absolute} rms squared.

\begin{equation}
    P_{k,\mathrm{abs rms}} = \frac{2 T}{N^2} |a_k|^2 \; ;
\end{equation}
In this normalization, the integral of the periodogram over a given frequency range will yield the squared rms in absolute units.

\begin{figure}
    \centering
    \includegraphics[width=\textwidth]{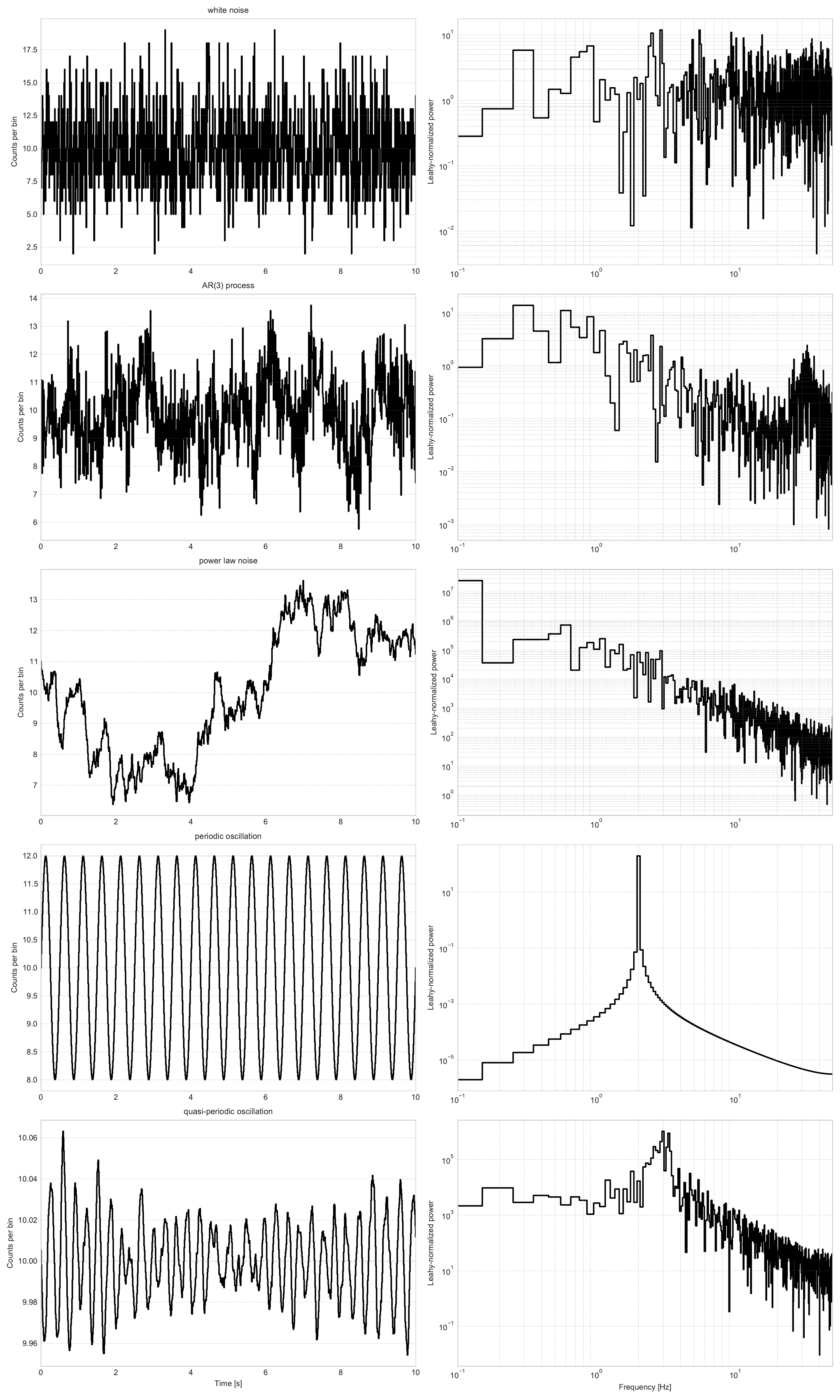}
    \caption{Different types of processes commonly encountered in X-ray sources. On the left, a we present a simulated light curve of the process. On the right, we show the corresponding periodogram. From the top: (1) white noise generated by a photon counting process. This corresponds to a source with a constant flux recorded by an X-ray detector; (2) an AR(3) process with parameters $C_1 = 0.2$, $C_2 = 0.1$ and $C_3 = 0.5$; (3) a red noise process with a power law spectrum with power law index $\rho_\mathrm{PL} = 2$ and a Gaussian flux distribution, simulated according to \citet{timmerGeneratingPowerLaw1995b}; (4) A strictly periodic, sinusoidal process with a period of $P = 0.5\mathrm{s}$, the only process in this selection that is not a stochastic one (though note how the presence of the window function smears out the delta function into a wider shape); (5) a QPO with varying signal amplitude and frequency, generated in the same way as (3).}
    \label{fig:proc}
\end{figure}

In all practical applications, the periodogram of observed data is not a completely faithful representation of the process that generated the data. For example, while in theory, a perfectly sinusoidal signal should produce a delta function at the frequencies corresponding to the signal's period $\nu_p$ and $-\nu_p$, this is not observed in practice (see also Fig \ref{fig:proc}). The reason lies in the assumption of the analytical (continuous) Fourier transform of a function of an infinitely long, continuous, stationary time series. In reality, however, observations are finite, and often data points are collected at discrete intervals. The multiplication of the original signal with a \textit{window function} and a \textit{sampling function} leads to a convolution between these three components in the Fourier domain (by virtue of the convolution theorem), which generally redistributes power and smears out the signal. More specifically, a rectangular window function (corresponding to turning on a detector at $t_\mathrm{start}$ and turning it off at $t_\mathrm{end} = t_\mathrm{start} + T$) will Fourier-transform to a sinc function, and thus the periodogram of a periodic time series subject to that window function will correspond to a convolution of that sinc function with the periodic signal's delta function. Similarly, the process of discrete sampling by a detector within an observation can often be well-described by a set of regularly spaced delta functions, also called a Dirac Comb. The Fourier Transform of this sampling function is also a Dirac Comb, reproducing aliases of the transform of the underlying process at intervals of $1/\Delta t$. This is one reason why extending the periodogram beyond the Nyquist frequency is discouraged, since signals appearing at frequencies higher than $\nu_\mathrm{Nyquist}$ are likely aliased versions of signals contained in the periodogram below $\nu_\mathrm{Nyquist}$.

Binning produces a loss of sensitivity when the frequency of the variable signal is not at the center of the spectral bin, and when it is close to the Nyquist frequency. From \citet{vaughanSearchesMillisecondPulsations1994}, the average response in a periodogram bin is:
\begin{equation}\label{eq:periodogramresponse}
    <P_j> = 0.773 N_{\rm ph} \frac{A^2}{2}\mathrm{sinc}^2\left(\frac{\pi}{2}\frac{\nu_j}{\nu_{\rm Nyq}}\right)
\end{equation}
where A is the fractional amplitude of the sinusoidal modulation.

\section{The Welch/Bartlett periodogram}
\citet{welchUseFastFourier1967} proposed a technique to smooth the periodogram by averaging N periodograms calculated from equal-length segments of the time series.
Assuming an evenly sampled time series $X(j), j=1\dots N$, one can extract from it $K$ sub-segments of length $L$, possibly overlapping:
\begin{eqnarray}
    X_1(j) =& X(j), j=1, \dots L \\
    X_2(j) =& X(j + M), j=1, \dots L \\
    X_3(j) =& X(j + 2M), j=1, \dots L \\
    (\dots)
\end{eqnarray}
with $M\leq L$. If $M=L$, segments are non-overlapping.
For each interval, one calculates the Fourier transform and the periodogram, as follows:
\begin{eqnarray}
    A_k(n) &=& \frac{1}{L}\sum^{L-1}_{j=0} X_k(j) W(j) e^{2kijn/L} \\
    P_k(n) &=& \frac{L}{\sum^{L-1}_{j=0} W(j)} {|A_k(n)|}^2
\end{eqnarray}
where $W(j)$ is a window function.
If $W$ is a rectangular window, there should be no overlap to maintain statistical independence of the time series bins.
In this case, the periodogram is also known as the Bartlett periodogram \citep{bartlettSmoothingPeriodogramsTimeSeries1948}.
Other choices of windows, that one might want to use to suppress side lobes
lead to the suppression of signal from certain bins, which should be recovered by a careful use of the hop distance $L$ in order to get a constant contribution from each bin to the final periodogram.

\section{Models for Commonly Encountered Signals}

In this section, we will give an overview of the kinds of signals commonly encountered in X-ray time series. None of these are physically motivated; rather, they are empirical descriptions of features observed in the periodogram. However, some of them have been successfully tied to physically meaningful interpretations of the phenomenology.

Of particular relevance to spectral timing is the type of process we assume to have generated the data. Most commonly known are \textit{deterministic} processes. These are processes for which future observations can be predicted exactly, given a functional form with known parameters. Examples of deterministic processes are a straight line, or a sine curve. They are to be distinguished from \textit{stochastic or random processes}, for which future observations cannot be predicted exactly. These are processes for which observations are random draws from a probability distribution.
A strictly deterministic process produces the same time series at times $t_i$ for the same parameters; a stochastic process does not. Instead, each \textit{realization} of a stochastic process can look quite different, even though the underlying physical mechanism might be the same.
A simple example of a stochastic process is one where observations at time $t_i$ are independent draws from a distribution, e.g. a Gaussian distribution such that $x_i \sim \mathcal{N}(\mu, \sigma^2)$, generating a noisy time series with variance $\sigma^2$ around a constant value $\mu$. One can formally distinguish between intrinsically stochastic processes, and those for which the underlying process is unknown (or unknowable) to a degree that they are impossible to predict. From a practical inference perspective, however, both might be well-modelled by a stochastic process.

A concept of particular importance in this context is \textit{stationarity}. A (strongly) stationary process is one whose \textit{joint probability distribution does not change as a function of time}. For example, a time series of independent observations drawn from a (multivariate) normal distribution is strongly stationary\footnote{note that independence is a much stronger requirement than stationarity, and the latter does not imply the former}. If one observes a realization from this process at time $t_i$ for $N$ time steps, and then later observes the same process again at time $t_j$ for $N$ steps, the two observed time series should be random draws from the same probability distribution. For \textit{weak stationarity} to hold, the joint probability distribution need not strictly remain constant with time, but the mean and variance of the process must be invariant with time. Weak stationarity is a concept that is a fundamental assumption to the statistical description of many of the methods described in this chapter, though we will also discuss issues arising from non-stationarity where appropriate.

In the following, we will highlight a number of different processes often considered in spectral timing. While they are often framed as features in the data, it is worth remembering that in practice, these are \textit{models} that describe an underlying (often unknown) physical process.

\subsubsection{Coherent signals}

We call coherent signals those characterized by a single fundamental and stable oscillation frequency, whose variation in period is very slow in time compared to the time needed to establish its existence such that we can assume stationarity over the interval it is being measured.
The most outstanding example of this stability are millisecond radio pulsars. These neutron stars can spin as fast as 765
rotations per second, sweeping the sky with radio beams like lighthouses.
Some of them are so stable that we can predict the arrival time of each pulsation with an uncertainty of $\sim$microseconds in years \citep{lorimerBinaryMillisecondPulsars2008},
making them excellent clocks comparable with the best atomic clocks.
Recent experiments like XPNAV-1 \citep{zhangMissionOverviewInitial2017} and SEXTANT \citep{witzeNASATestProves2018} are trying to use pulsars as a navigation system, determining the position of spacecraft by using the apparent frequency of pulsars as a reference.
Another example of coherent signals are stellar pulsations, driven by the periodic expansion and contractions of the outer envelopes of stars. In some cases, this pulsation is very regular, for example in the class of variable known as regular Cepheids \citep{madoreCepheidDistanceScale1991}.

\subsubsection{Stochastic Processes}
\label{sec:stochproc}
Many astrophysical sources, such as X-ray binaries, show brightness variations at high energies that are not easily modelled with a deterministic process but that can be expressed well in terms of stochastic processes. We will briefly highlight two such models, but refer the reader to \citet{feigelsonTimeDomainMethods2022a} and to \citet{scargle1981studies} for more details.

As mentioned above, stochastic processes are those for which future observations cannot be predicted exactly, but where observations involve draws from some probability distribution. One of the simplest and most common stochastic process is a \textit{white noise process}. This process involves uncorrelated (though often enough it is reasonable to assume independent) draws from a probability distribution. Like the other major class of stochastic processes introduced later in this section, \textit{red noise}, it is named in analogy to the spectrum of visible light: white noise, by definition, has the same expected underlying power at all temporal frequencies. A typical example might be random noise from a detector. Here, the detector is often assumed to generate independent measurements in each time bin, and thus it can be reasonable to assume that the noise in bin $i$ does not depend upon the noise in bin $i-1$ (though there are examples where this assumption is not fulfilled, e.g. \citealt{2013arXiv1309.1176W}). in X-rays, where telescopes record individual photon arrival times which are subsequently binned into light curves measuring counts as a function of time. These counts are usually well-described by a Poisson distribution with a rate parameter given by the brightness of the underlying source, and if the detectors behave well, each time bin will consist of an independent draw from this distribution. However, there are effects that can break this assumption (see Section on \textit{dead time}) and which significantly alter the statistical description and breaks the assumption of independence.

Uncorrelated white noise is a a standard null hypothesis of many periodicity detection methods, as we will see later, but it is not a good description of variability in many astrophysical sources. XRBs and AGN often show light curves that vary on longer timescales than a white noise process can describe. A subset of these kinds of stochastic process are often also called \textit{red noise} processes---again in analogy to the visible spectrum---and have higher-amplitude variability at low frequencies compared to high frequencies (thus being ``red'').

There are many models that can describe physical systems with variability on multiple time scales, but some of the most useful are the  the \textit{autoregressive} and the \textit{moving average} model. As an example of an autoregressive process, consider a coin toss experiment, where a coin is tossed $N$ times, and for each instance of ``heads'', an increment of $+1$ is added to a running sum, whereas for ``tails'', $-1$ is subtracted from the running sum. Consider the running sum as a function of time as the variable of interest: for any $t_i$, the value of $x_i$ depends directly on a random draw from a probability distribution (for a coin toss the binomial distribution) as well as the directly preceding value, $x_{i-1}$. This is the fundamental setup for an autoregressive process of order 1.

More formally, one can write for an autoregressive process of order q, also denoted as $AR(q)$:

\begin{equation}
    x_i = R_i + \sum_{k=1}^{q}{C_k x_{i-k}} \; .
\end{equation}

Here $R_i$ is a draw from a random distribution, also called the \textit{innovation} because it injects new information into the process. The second term describes how the current observation depends on the $q$ previous one, given coefficients $C_k$.

The moving average (MA) process is closely related to the AR process, and is of particular interest for spectral timing. MA processes involve a convolution of a random draw (an innovation) with an \textit{impulse response function}, which describes how the future output $x_i$ depends on the convolution of previous innovations and the impulse response function. Examples are, for example, the response of a piece of equipment to an incoming system, but also the response of gas in an accretion disk at a given radius toward incoming radiation from close to the black hole. More formally, the MA process consisting of an innovation of uncorrelated white noise $R$ and an impulse response function $C_k$ can be written as

\begin{equation}
    x_i = \sum_{k=-\infty}^{\infty}{C_k R_{i-k}} \; .
\end{equation}

The autoregressive process models how a current observation depends on a set of previous observations, while the moving average process models how a system \textit{responds} (via the impulse response function) to a random variable. Note that the impulse response function is only \textit{causal} if $C_k = 0$ for $k < 0$. While this should generally be true in problems set in the time domain, both AR and MA processes are useful in other contexts as well, where they may not be as strict a causal requirement as there is in the time domain.

\subsubsection{Quasi-periodic signals}

Many accreting systems show some kind of quasi-periodic oscillations (QPO) in their emission. These are (almost) periodic signals where the frequency of the oscillation or the amplitude of the oscillation (or both) vary as a function of time, and where the amplitude and/or frequency can themselves be described by a stochastic process.

The model of choice in the Fourier domain for QPOs is often a Lorentzian function (in its form as a probability density known as the Cauchy distribution), defined as \citep{nowak2000}

\begin{equation}
    \label{eqn:lorentzian}
    f(\nu) = \frac{A}{\gamma\pi} \frac{\gamma^2}{(\nu - \nu_c)^2 + \gamma^2} \; ,
\end{equation}

\noindent with amplitude $A$, $\gamma$ the half width at half maximum, and $\nu_c$ the centroid frequency. The \textit{quality factor} or $q$-factor of a QPO is defined as the ratio of centroid frequency to full width at half maximum, $q = \nu_c / 2\gamma$. In many applications, it is conventional to not allow a QPO to be arbitrarily broad (spread over enough frequencies, as QPO becomes essentially indistinguishable from an aperiodic stochastic process), but define that $q \geq 2$ for the existence of a QPO \citep{vanderklisRapidXrayVariability2006}. Similarly, QPOs with large $q$-factors in the range of $50-100$ tend to concentrate their power in a single Fourier frequency bin, and become indistinguishable from strictly periodic signals.

The richest phenomenology is found in accreting neutron stars, where QPOs are observed over a wide range of frequencies from milli-Hz to kHz  \citep{vanderklisOverviewQPOsNeutronstar2006}. These frequencies are generally correlated with one another, implying a common physical origin, such as the Keplerian frequencies at specific radii of a disk approaching the NS.
Black hole binaries also show various kinds of QPOs, with frequencies ranging from milli-Hz to hundreds of Hz. BH QPOs are often modeled with various kinds of relativistic phenomena \citep{stellaLenseThirringPrecessionQuasiperiodic1998}. The interpretation of NS QPOs is more complicated, having the additional component coming from the strong magnetic field of the NS \citep{vanderklisOverviewQPOsNeutronstar2006}.
The single oscillations of QPOs are rarely detectable, but when they are, they show extremely interesting behavior. One spectacular example is the source GRS 1915+105, showing a rich collection of impulse shapes, from sinusoids to ``heart beats’’.
Recently, some AGNs have been shown to show heart beats similar to GRS 1915+105, a phenomenon that has been called Quasi-periodic Eruptions \citep{miniuttiNinehourXrayQuasiperiodic2019}.

\subsubsection{Fast Transients}

Transients are, by definition, \textit{non-stationary} phenomena. They are described by a rapid, time-bounded variation in brightness, and can be well-described by deterministic functions (e.g.~supernovae) or by stochastic processes (e.g.~magnetar bursts, \citealt{huppenkothen2013}). Depending on the type of source, transients may be the only physically relevant signal in the data (i.e.~the only other source of variation is the measurement process in the detector; examples are gamma-ray bursts and supernovae) or they may occur on top of other brightness variation created by the same source (e.g.~stellar flares, flares in accreting supermassive black holes like Sgr A*). Detecting flares and other transients is an important, but very broad problem that will be discussed elsewhere. For this chapter, the key property of transients to keep in mind is their lack of stationarity \textit{compared to the time scales of interest}. The time spans at which one intends to probe physically meaningful timescales are of critical importance here. BHXRBs show non-stationary behaviour on timescales of weeks to months in the form of outbursts. However, for studies interested in rapid variability on timescales of ~seconds or shorter, individual observations of thousands or tens of thousands of seconds may be considered stationary for the purposes of generating and analysing Fourier products (even though the properties of rapid variability also changes throughout an outburst).
As we will see towards the end of the Chapter, non-stationarity on timescales relevant to the scientific question at hand can significantly complicate the characterization of variability and the detection of (quasi-)periodic oscillations in these sources, and mitigation strategies must be carefully considered.

\section{Periodogram Statistics}

The statistics of random processes is generally well known. Under the conditions encountered in X-ray astronomy (evenly spaced, Poisson-distributed time series) it is generally reasonable to assume that the complex Fourier amplitudes are Gaussian random variables with a zero mean and a variance that is set by the underlying power spectrum that is believed to have generated the data. Consequently, the periodogram, defined as the square of the of the absolute value of the Fourier amplitudes, will follow the distribution of the sum of squares of the real and imaginary (Gaussian) Fourier amplitude components. The latter is well-known: the square of a Gaussian random variable is a $\chi^2$ random variable with one degree of freedom. The sum of two $\chi^2_k$ random variables (with $k$ degrees of freedom) is also a $\chi^2$ random variable with $k+k=2k$ degrees of freedom. For white noise (e.g. photon counting statistics without variability in the time series), all powers in the periodogram are independently distributed around a common mean. The Leahy normalization is defined such that the periodogram is distributed according to the standard, central $\chi^2_2$ distribution, i.e. centred on a mean of $2$ and with a variance of $4$. For a stochastic process, the periodogram follows a $\chi^2_2$ distribution, but centred on the true underlying power.

The periodogram is an \textit{inconsistent} estimator of the underlying power spectrum. That is, there is an intrinsic variance in the periodogram of a stochastic process that is \textit{not} reduced by adding more data points. This is because adding more data points will add more \textit{frequencies} to the periodogram, but not change the statistical distributions of each individual power in the periodogram. Taking a longer observation corresponds to an extension of the periodogram to lower frequencies, because the longer baseline makes it possible to add sinusoids with longer periods. Conversely, increasing the sampling rate enables looking at higher frequencies, thus adding more data points to the high-frequency end of the periodogram by shifting up the Nyquist frequency.

One \textit{can} improve the performance of the periodogram estimator through averaging, either of neighbouring frequencies in the periodogram, or of periodograms generated from segmenting the observation into individual, smaller time series. Because sums of $\chi^2$ distributed variables are also $\chi^2$ distributed variables, the average of $L$ periodograms of independent segments yields a $\chi^2_{2L}/2L$ variable, with a smaller variance than the original periodogram (the division by $2L$ comes from the fact that we generally \textit{average} segment periodograms rather than simply sum them). Similarly, averaging $M$ neighbouring frequency bins yields a $\chi^2_{2M}/2M$ distributed variable around the true power spectrum. Note that one can (and in practice often does) combine both averaging of frequencies and of segment periodograms.

\subsection{The Likelihood for Periodograms}

The true underlying power spectrum is often unknown in practical applications, and might be a quantity of interest to be estimated. For example, one might be interested in the centroid frequency of a QPO, and how it changes throughout the outburst of a BHXRB. While one in principle can estimate that centroid frequency from the periodogram directly by finding the largest power in a frequency range encompassing the QPO, the latter is a (potentially very) noisy estimator. Fitting a model to the periodogram may yield a more reliable estimate of the centroid frequency, because it \textit{jointly} models a set of powers assumed to all belong to the same power spectral feature. This requires a well-chosen model for the underlying power spectrum, of course. As we have seen earlier QPOs are usually modeled with a Lorentzian function, whereas for red noise, power laws and broken power laws are often used as a model.

The $\chi^2$ distribution introduced above defines a \textit{sampling distribution} or \textit{likelihood}, i.e.~the probability distribution of observing a given data set (here, we will consider the set of periodogram powers as our data set) given some underlying (true, unknown) power spectrum. If we define a model power $S_j(\theta)$ at frequency $\nu_j$, specified by a set of parameters $\theta$ (for example the QPO amplitude, centroid frequency and full width at half maximum), we can then compute the probability of having observed periodogram power $P_j$ at that same frequency \citep{vaughanBayesianTestPeriodic2010},

\[
p(P_j | S_j(\theta)) = \frac{1}{S_j(\theta)} \exp{(-P_j / S_j(\theta))} \; .
\]

\noindent Note that for now, we will assume that the periodogram has not been averaged across segments or neighbouring frequency bins for the equation above to hold.

The likelihood for a periodogram over $N/2$ observed powers $P_j$ is then defined as the product of individual probabilities for each frequency $\nu_j$. One generally (and equivalently), defines the logarithm of the likelihood as the sum of logarithm of all probabilities, such that

\begin{eqnarray}
\label{eqn:psdloglike}
    \log(\mathcal{L}(\theta)) & = & \sum_{j=1}^{N/2} \log(p(P_j | S_j(\theta))) \nonumber \\
                            & = & \sum_{j=1}^{N/2} \left(\log(S_j(\theta))-\frac{P_j}{S_j(\theta)}  \right) \; .
\end{eqnarray}

\noindent A maximum likelihood solutions for the unknown parameters $\theta$ can be found by minimizing the negative log-likelihood,
\[
\theta_\mathrm{max} = \mathrm{argmin}_\theta{(-\mathcal{L}(\theta))} \; ;
\]

\noindent with common numerical optimization routines.
Alternatively, \citet{vaughanBayesianTestPeriodic2010} proposed adding prior probability densities for the parameters $\theta$ to the problem, and sampling the posterior probability density via Markov Chain Monte Carlo.

A likelihood for averaged periodograms was derived from the $\chi^2_{2LM}/2ML$ sampling distribution for periodograms averaged over $L$ independent segments and $M$ independent neighbouring frequencies by \citet{barretMaximumLikelihoodFitting2012}:

\begin{equation}
\label{eqn:avgpsdloglike}
    \mathcal{L}_\mathrm{avg}(\theta) = -2ML \sum_{j=1}^{N/2} \left\{ \frac{P_j}{S_j(\theta)} + \ln{S_j(\theta) + \left( \frac{1}{ML} - 1 \right)\ln{P_j} + c(2ML) }\right\} \; ,
\end{equation}

where $c(2ML)$ is a factor independent of $P_j$ or $S_j$, and thus unimportant to the parameter estimation problem considered here (it only scales the likelihood, but does not change its shape).

These likelihoods form the basis for a range of possible analyses, from estimating the fractional rms amplitude in a feature to QPO detections (considered further down below in this Chapter).
\begin{figure}
    \centering
    \begin{subfigure}[b]{0.55\textwidth}

    \includegraphics[width=\textwidth]{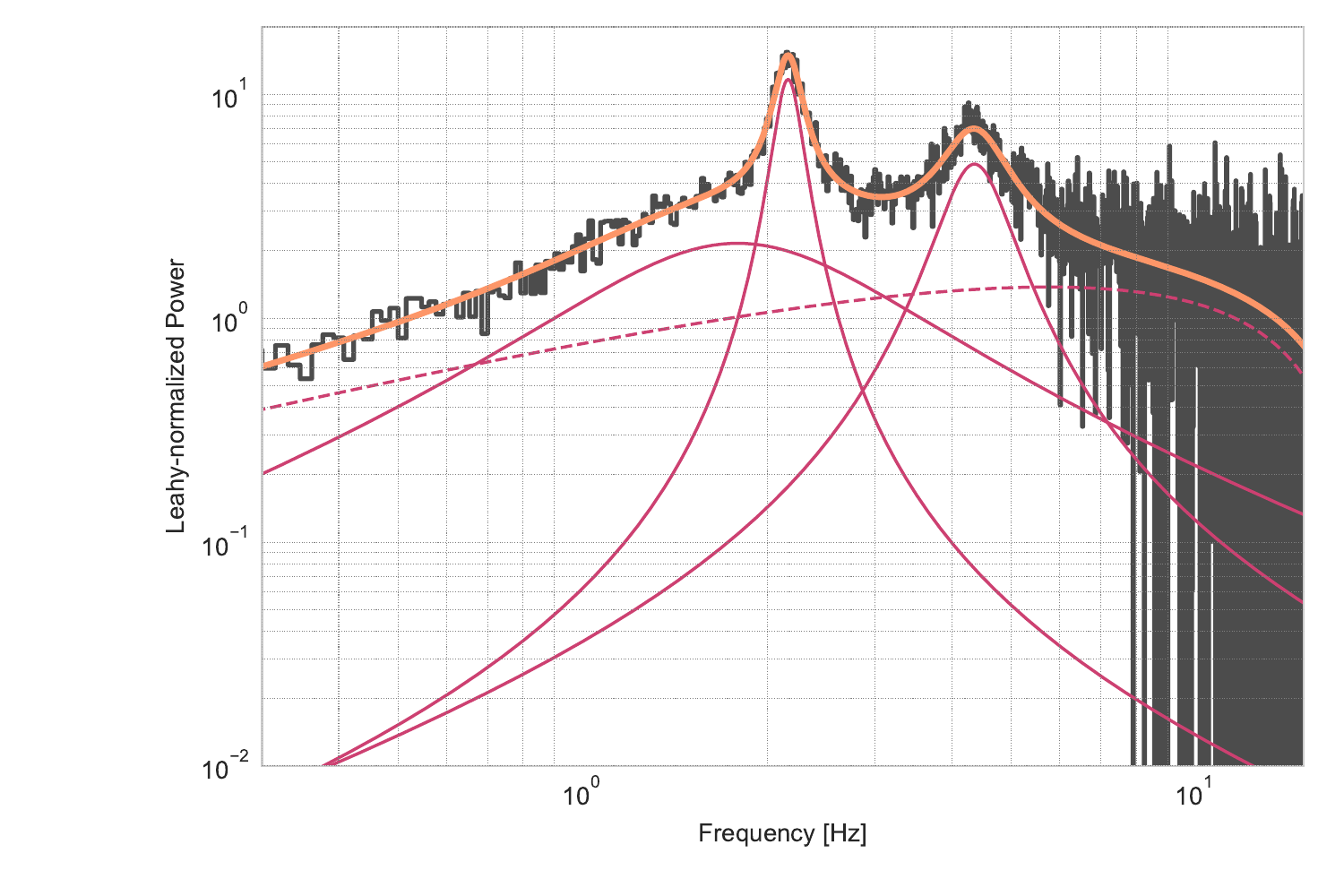}
    \end{subfigure}
    \hfill
    \begin{subfigure}[b]{0.4\textwidth}
    \includegraphics[width=\textwidth]{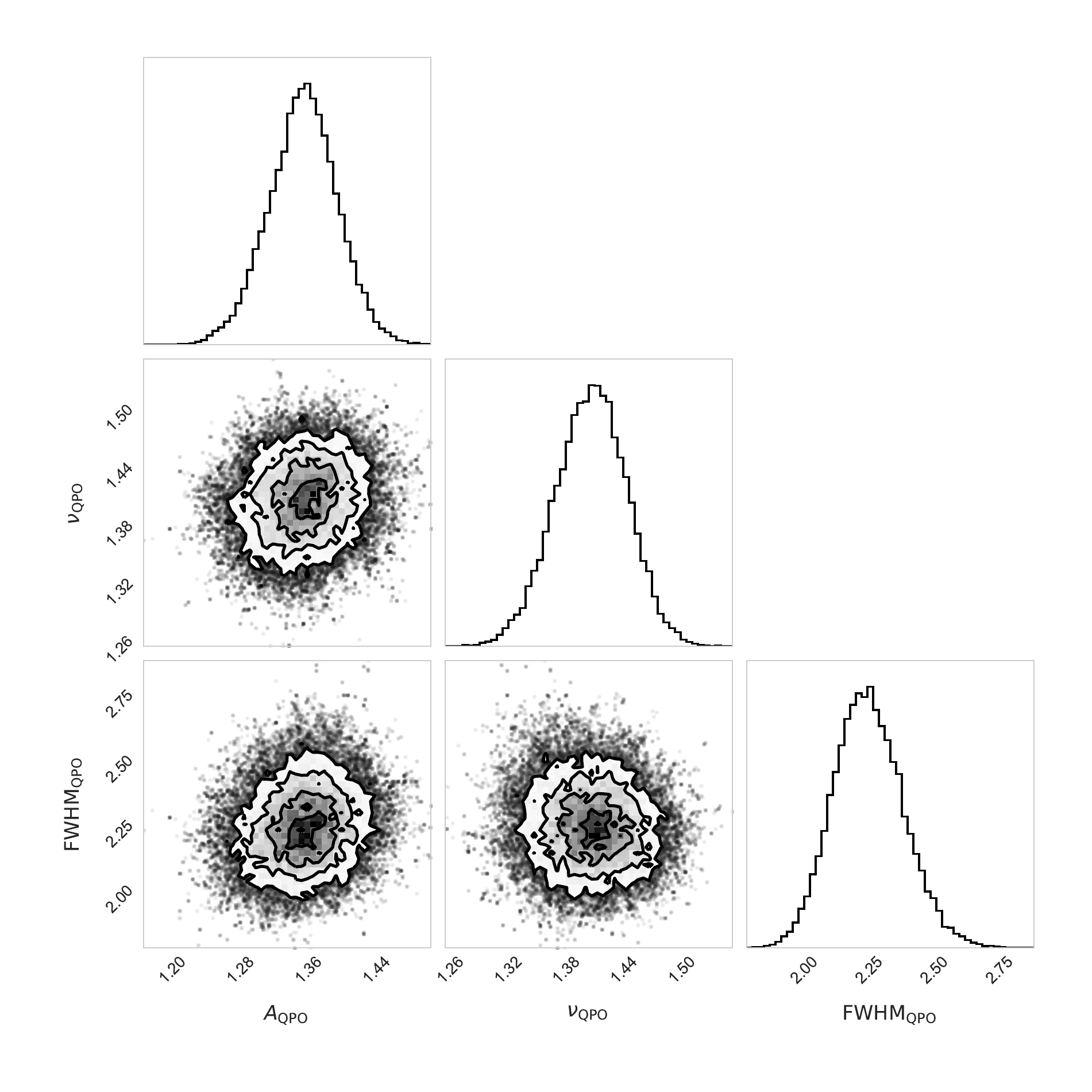}
    \end{subfigure}
    \caption{
    Left: Periodogram of a NICER data set for the galactic black hole binary GRS 1915+105 (gray) and a maximum likelihood fit (orange) to the periodogram of a model with a power law component for low-frequency stochastic (aperiodic) noise and three QPO components parametrized as Lorentzians (individual model components in purple). The optimization used a negative version of the log-likelihood defined in Equation \ref{eqn:avgpsdloglike} to find the maximum likelihood parameters. Right: Posterior distributions for the QPO parameters sampled using the \texttt{emcee} package \citep{foreman-mackeyEmceeMCMCHammer2013} using a combination of the likelihood defined in Equation \ref{eqn:avgpsdloglike} and flat, uninformative priors.}
    \label{fig:grs1915_maxlike}
\end{figure}

\subsection{Simulating Stochastic Time Series}
\begin{figure}
    \centering
    \includegraphics[width=\linewidth]{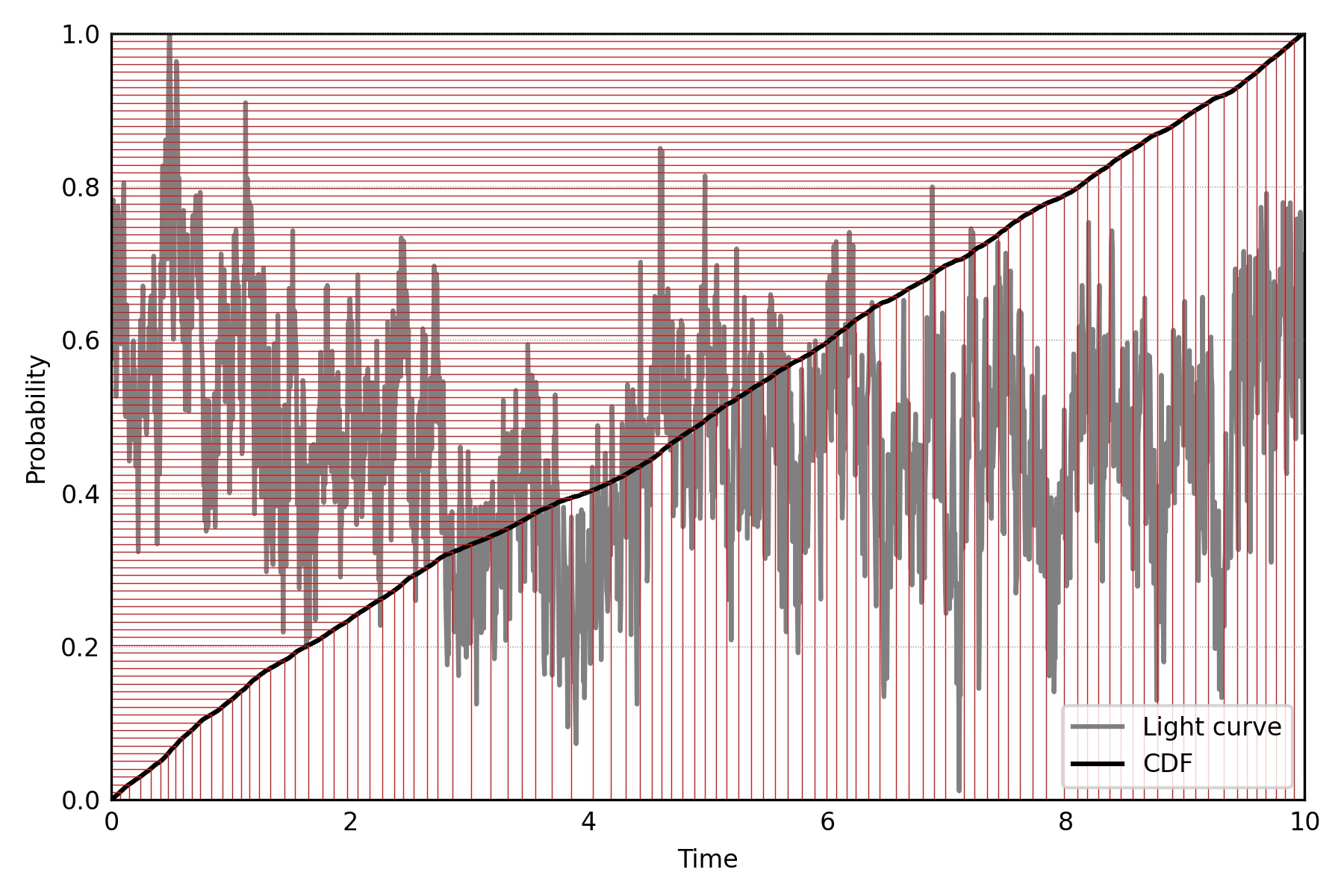}
    \caption{The inverse CDF method to simulate event lists from continuous light curves.
    Given a positive-definite light curve (generated, e.g., with the method by \citealt{timmerGeneratingPowerLaw1995b}), we treat it as a probability distribution: we calculate the cumulative distribution function by calculating its cumulative sum and normalizing to 1. Then, we generate random numbers uniformly distributed between 0 and 1 (horizontal lines) and take the event times at the corresponding values of the CDF (vertical lines).}
    \label{fig:CDF}
\end{figure}

In many applications, it is useful to generate time series from a power spectral shape and a stochastic process. Ideally, this generates simulations that look similar to some observed data, and can often help understand and mitigate instrumental effects, or enable estimation of accurate systematic and random uncertainties. This process starts with assuming a parametric model for the underlying power spectrum that generated the data in question, usually in the Fourier domain (For example the above-mentioned Lorentzian for a QPO or a power law to simulate red noise). Given a set of parameters (for example estimated via maximum likelihood estimation on the data), the value of that parametric model at a range of frequencies will provide the underlying power of the process at those frequencies. However, we are often looking for one or multiple \textit{realizations} of that process, which requires a form of randomization. The most common version of this is based on the algorithm by \citet{timmerGeneratingPowerLaw1995b}, which takes advantage of the statistical properties of the periodogram and generates two Gaussian random variables for each frequency (for the real and imaginary components of the Fourier amplitude, respectively) with a mean of zero and a variance that depends on the power at that frequency (the exact formulation of the variance depends on the normalization of the power spectrum). These Gaussian random numbers can be turned into a time series through an inverse Fourier transform. Because this produces strictly Gaussian flux distributions in the resulting time series \citet{emmanoulopoulosGeneratingArtificialLight2013} extended this framework in order to reproduce both a power spectral shape and a non-Gaussian flux distribution.

Alternatively, starting from a Gaussian flux distribution one can simulate Poisson-distributed light curves in multiple ways.
The most straightforward is the direct generation of Poisson random numbers using the flux simulated above as the local mean in each light curve bin. This produces light curves with more realistic statistical properties for an X-ray signal.
Going further, one can simulate a list of events using an inverse cumulative distribution function (CDF) method: using the light curve as a probability distribution, calculate numerically its CDF; then, simulate $N$ values of probability $p_i$ uniformly distributed between 0 and 1; finally, and record the times $t_i$ whose CDF values are equal to $p_i$ (See Figure~\ref{fig:CDF}).

\begin{figure}
    \centering
    \includegraphics[width=\textwidth]{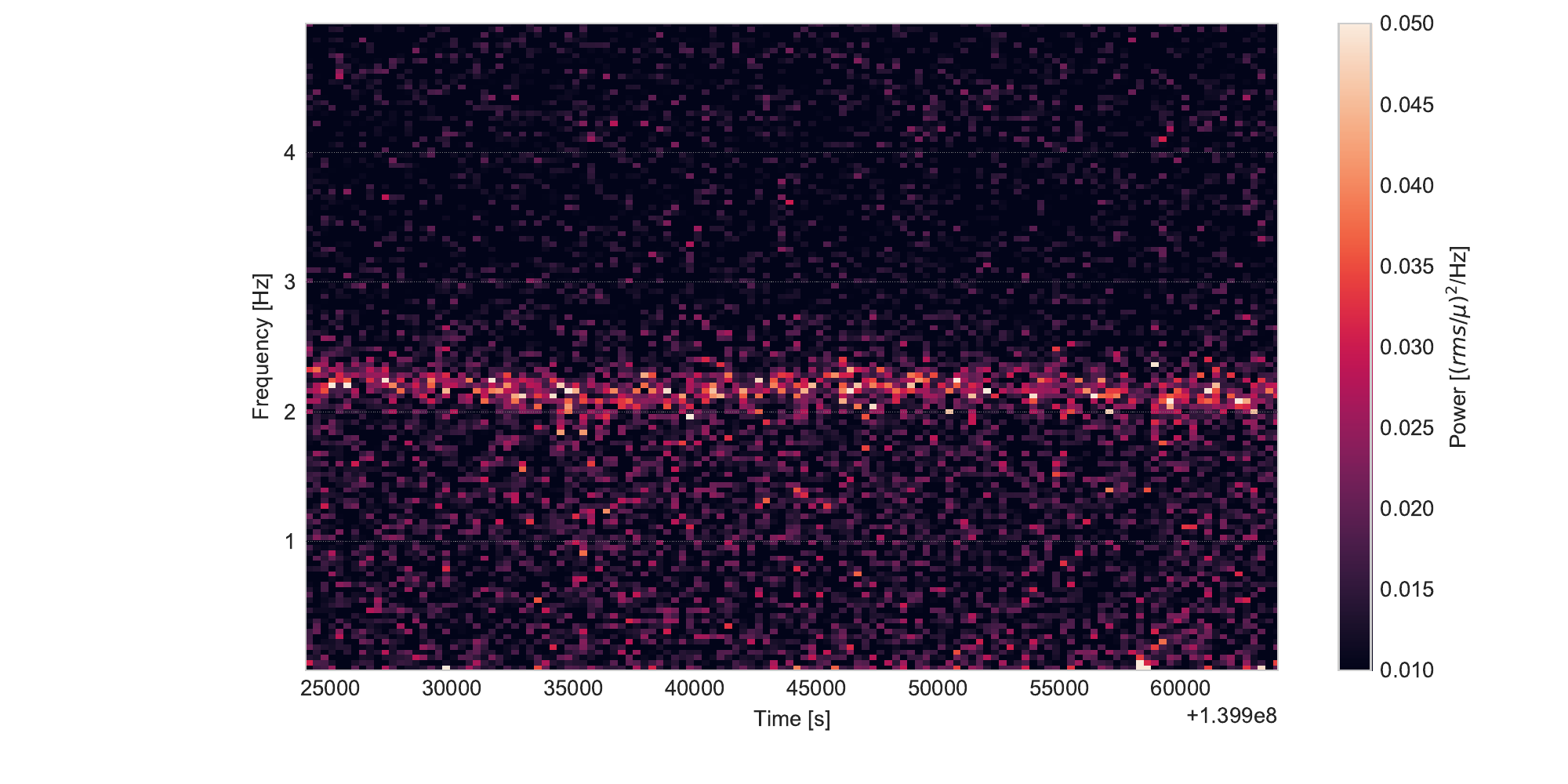}
    \caption{Dynamical periodogram generated from the same NICER data of GRS 1915+105 as shown in Figure \ref{fig:grs1915_maxlike}. The strong QPO at $\sim 2.5\,\mathrm{Hz}$ is clearly visible as a bright band throughout the observation, and there appear to be potential modulations to the centroid frequency, though this has not been confirmed through a more thorough analysis. Note that data gaps in the observation have been removed to improve clarity of the figure.}
    \label{fig:dynspec}
\end{figure}

\subsection{Dynamical periodograms}

When defining QPOs earlier, we asserted that for these signals, the signal's oscillation frequency might not be static. In many accreting sources, it has been known to jitter around a central value. In other sources, QPOs can drift in frequency over the course of a long observation, or might vanish for an interval before reappearing. Generating a single periodogram from such an observation does not allow probing the exact behaviour of a QPO as a function of time. One may segment an observation into a number of shorter intervals, and then generate a periodogram for each one. This allows for the representation of the data as a \textit{dynamical power spectrum} (which, in most fields outside of astronomy, is primarily called a \textit{spectrogram}) through plotting a map of the power as a function of time and frequency (for an example, see Figure \ref{fig:dynspec}). If the signal is of sufficiently large amplitude and frequency to be reliably detected in each segment (should it be present), this representation allows for a straightforward visual trace of the QPO's behaviour of the course of an observation. For a more detailed introduction in the context of thermonuclear burst oscillations, see e.g.~\citep{watts2012}.

\section{Periodicity Detection}
\label{sec:periods}

Periodic and quasi-periodic signals are of particular interest in many astrophysical applications. They represent a process repeating on a potentially very regular timescale: this often constrains relevant physical processes, and the physical extent of the emitting region. For example, in pulsars, the highly periodic radio emission is the astrophysical signature of the star’s rotation, a fundamental quantity in constraining formation models.

At first glance, detecting a periodic signal might appear straightforward: periodic signals are highly regular patterns, especially when they are also sinusoidal in shape, that are easy to recognise by eye. In practice, however, constraining periodic signals is only easy in the most obvious of cases. For example, a signal may be periodic but not sinusoidal: instead, perhaps it consists of regular short bursts of emission. This tends to spread signal power over multiple harmonic frequencies in the Fourier domain. Or an observation might be very faint, such that the stochastic nature of the instrument noise renders the signal invisible in the light curve, and makes it difficult to tell whether a peak in the periodogram may simply be due to the instrumental noise, rather than a signal. Quasi-periodic signals have, by their nature, power that is spread across multiple neighbouring frequencies, such that detection methods should take this structure into account. An observed light curve might be composed not only of a periodic signal, but multiple such signals, along with non-periodic brightness variations that all must be disentangled.

As a response to these challenges, various fields of astronomy have developed a wide arsenal of methods to detect and characterise periodic and quasi-periodic signals in light curves. Here, we will primarily focus on Fourier-based methods of detection, but will highlight other approaches where relevant. There are a number of questions a researcher may want to ask themselves when attempting to detect a (quasi-)periodic signal. The responses to these questions crucially determine which methods will be appropriate for detection and characterisation of the putative signal.

\begin{itemize}
\item{\textbf{Do we know that a periodic signal exists, or do I just suspect it and am trying to find it?} Characterization of a signal known to exist is generally simpler than (blind) searches for potential signals in data. Thus, the characterization problem often allows for a simpler analysis.}
\item{\textbf{Is the signal expected to be periodic or quasi-periodic?} Strictly periodic processes tend to be easier to detect through folding or Fourier methods. In Fourier space, these signals generate a very narrow peak, often confined to a single frequency bin, whereas QPOs do not.}
\item{\textbf{Is the light curve evenly sampled on the timescales of interest?} Temporal structure in the sampling pattern (e.g. through data gaps) can confuse certain period detection methods, especially those based on the Fourier transform. If the signal to be detected has a much shorter period compared to a single, regularly sampled observing window, then this might not matter. However, if the signal is of comparable period as the observing pattern, one should employ methods that take sampling patterns into account. }
\item{\textbf{What other variability do I expect from the source?} Periodicity detection is easiest when the signal in question is the only signal expected in the data, and the source is otherwise at constant flux. Broadband variability, for example, in the form of red noise, significantly complicates detection processes, and spells trouble for any visually-based approaches to detection, because our human ability of pattern recognition plays tricks on our brain. }
\item{\textbf{Is the light curve stationary or non-stationary?} When the statistical properties of the light curve change as a function of time on the timescales we aim to generate Fourier products on, this can impact the validity of various period finding algorithms, which often assume stationarity. This is particularly relevant when searching for (quasi-)periodicities in fast transients like GRBs, magnetar bursts and FRBs.}
\end{itemize}

Let us now introduce a range of available, primarily Fourier-based detection methods in order of complexity.

\subsection{Signal Detection in Constant Noise}

The simplest case is that of a stationary, strictly periodic signal in an infinitely long, evenly sampled light curve consisting only of a constant background and detector noise (i.e.~white noise), because the Fourier transform of a constant is a Delta function centered on zero, and so the theoretical expectation of the background spectrum is well-defined and can be assumed to be known.

\begin{figure}
    \centering
    \includegraphics[width=\textwidth]{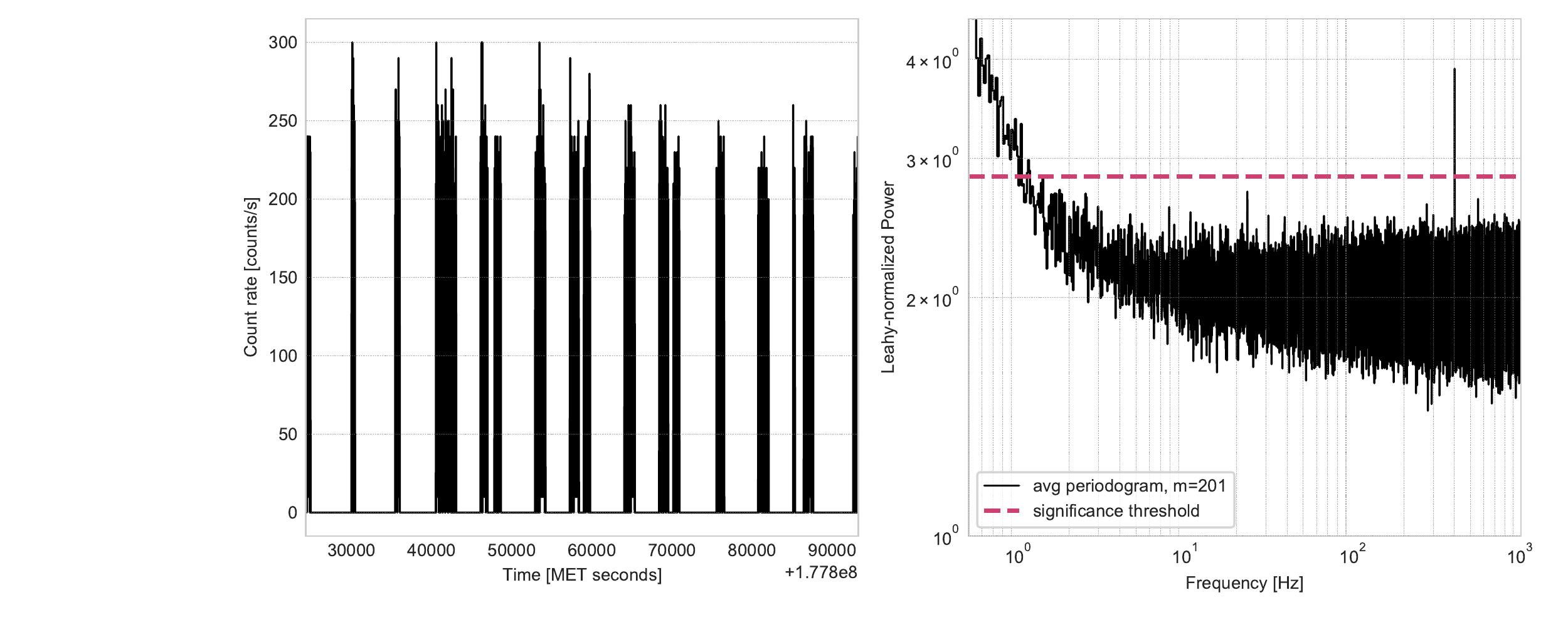}
    \caption{Light curve (left) and periodogram (right) of the neutron star X-ray binary SAX J1808.4-3658 observed with NICER. Here we show an averaged periodogram generated out of 201 segments of 64 seconds duration each. The neutron star's spin frequency is clearly visible as a narrow peak at 401 Hz. The dashed line identifies the threshold of a $p=0.003$ ($\sim3\sigma$) detection, corrected for the $65535$ frequencies in the periodogram (assuming a blind search for an signal of unknown frequency). At low frequencies, a common failure mode of this analysis becomes apparent: a red noise component leads to large variance (and consequently a large periodogram power) at low frequencies, yielding ``significant'' detections for all frequencies below $\sim 1.2\,\mathrm{Hz}$. However, these do not all represent periodic signals. Rather they represent a frequency band where the core assumption of the model (white noise) is not accurate.}
    \label{fig:class_sig}
\end{figure}

Based on the statistical properties of the periodogram defined in the section on periodogram statistics, we can define a statistical test using the $\chi^2$ distribution. This test is based on the null hypothesis that there is \textit{no signal} in the light curve (for an introduction into statistical tests using null hypotheses, see e.g.~\citealt{vaughan2013scientific}). Consequently, all powers should be consistent with random draws from a $\chi^2_2$ distribution for a single periodogram, or a $\chi^2_{2ML}/2ML$ distribution for averaged periodograms. A power that is significantly higher than expected under that distribution is often taken as evidence that a periodic signal exists in the data. In practice, one computes a tail probability (also called a $p$-value) by integrating the probability above the observed power of interest. If this probability is below a specified value (e.g.~$p<0.003$, roughly corresponding to a 3$\sigma$ detection), one assumes that this power is unlikely to be generated by white noise only (for an example, see Figure \ref{fig:class_sig}). There are a few important caveats to this procedure. Most importantly, note that this is a test that \textit{rejects} a null hypothesis, rather than confirms the presence of a periodic signal. That is, one may also detect ``significant’’ periodic signals if none are in the data, but the assumptions of the null hypothesis (even sampling, constant light curve, Gaussian Fourier amplitudes) are broken. This can for example happen when the source in question is not strictly constant, but contains aperiodic variability. As we will see later, certain detector effects like \textit{dead time} can change the statistical properties of the periodogram and render this test invalid.

Whereas strictly periodic oscillations will generally be contained in a single frequency bin, the power in quasi-periodic oscillations is spread across multiple frequencies. In this case, outlier detection on a single frequency bin may not yield a significant detection, but adding multiple neighbouring bins can add in enough extra power to lift the signal above the detection limit, especially when taking into account the increased detection power from the $\chi^2$ distribution for averaged bins.
Averaging powers, whether it is by averaging neighbouring frequencies or periodograms of independent segments, decreases the statistical variance in the periodogram. As a consequence, averaged periodograms are potentially more sensitive to aperiodic and quasi-periodic signals than non-averaged periodograms, if the signal to be found is stationary. Which strategy--averaging neighbouring frequency bins, averaging segments of light curves, or both--is chosen in practice depends on the expected signal in question. If the expected signal has a very short period (i.e.~a high frequency) compared to the length of the observation, then averaging the periodograms of short segments may be a useful method to boost signal-to-noise ratio. If the signal is expected to be at low frequencies, however, averaging short segments may be less beneficial, because each segment might only contain a very few cycles (or even just one) of the signal.
The choice might also be affected by observing conditions, e.g., the maximum time frame for periodogram segments might be limited by the periodic occultations from the Earth during the satellite's orbital motion.

\begin{figure}
    \centering
    \includegraphics[width=\textwidth]{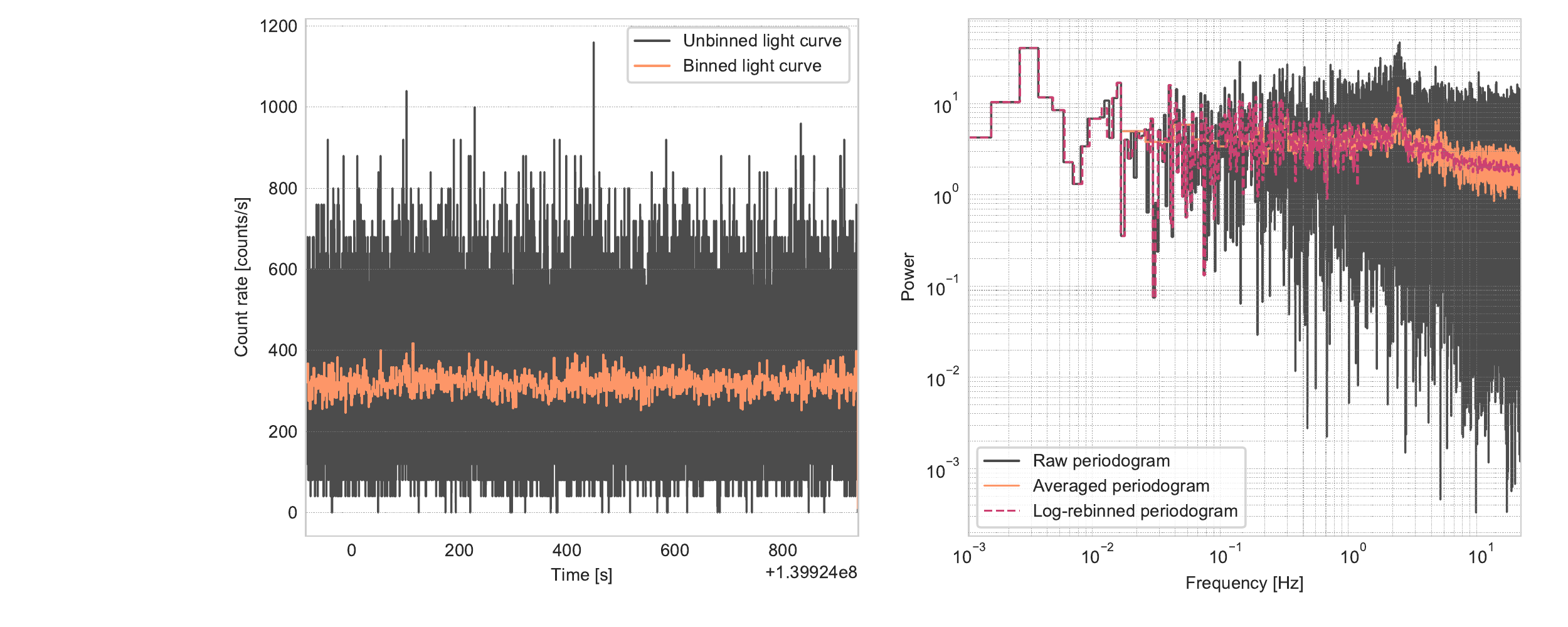}
    \caption{A single Good Time Interval (GTI) of the NICER observation of GRS 1915+105 introduced in Figure \ref{fig:grs1915_maxlike}. Left: light curve with two different time resolutions. Right: periodogram and two different periodogram binning schemes. In dark grey, we show the unbinned periodogram. The logarithmically binned periodogram in dashed purple starts out equivalent to the unbinned periodogram, but bins become wider with increasing frequency and the noise distribution narrows as more and more powers are averaged per bin. In solid orange, finally, we show an averaged periodogram generated out of 15 individual segments. Because of the smaller duration of each segment compared to the unbinned periodogram, the averaged version starts at a higher frequency and has a narrower, but constant noise distribution throughout.}
    \label{fig:binning}
\end{figure}

For searches across a multiple orders of magnitude in Fourier frequencies, a logarithmic binning scheme (where bins become progressively larger at higher frequencies) can be helpful in QPO discovery, because a QPO at large frequencies will span a wider range of frequencies as a QPO at smaller frequencies with the same quality factor. Note, however, that the detection sensitivity becomes frequency-dependent in this case. For an example of different binning schemes, see Figure \ref{fig:binning}.

For blind searches (i.e.~searches where the signal's frequency is not known in advance and a search is conducted across all available frequencies in the periodogram or a subset of frequencies) it is important to correct the significant threshold by the \textit{number of trials} (also called the \textit{Bonferroni correction}), i.e.~ the number of independent frequencies and periodograms searched. This is because when searching a periodogram with $1000$ frequencies, a power with a p-value $p<0.001$ will, by definition, occur purely by chance on average once.

\subsection{Upper limits on the pulsed amplitude}
When comparing the results of searches over multiple parameters or calculating confidence intervals, the null hypothesis is not anymore the absence of signal, but that there is a signal with a given power.
Therefore, we use an approach similar to \cite{brazierConfidenceIntervalsRayleigh1994} to evaluate the confidence intervals, starting from the following formula for the probability to detect a signal with the Rayleigh test.
Given a signal

\begin{equation}\label{eq:sinus}
    y = \lambda(1 + a \sin{\omega t}),
\end{equation}
the expected power in a single periodogram in \citet{leahySearchesPulsedEmission1983} normalization or in the $Z^2_1$ statistic (see below) calculated at the correct frequency from $N$ photons is
\begin{equation}\label{eq:psig}
\psig=N\frac{a^2}{2}.
\end{equation}
The probability to measure a given power $\pmeas$ is then given by \citep{brazierConfidenceIntervalsRayleigh1994,vaughanSearchesMillisecondPulsations1994}:

\begin{equation}\label{eq:xsq}
p(\pmeas | \psig) = \mathrm{e}^{-(\pmeas+\psig)/2}\,I_0\left(\sqrt{\pmeas\psig}\right)
\end{equation}

where $I_0$ is the modified first-order Bessel function of order 0.
Note that there is a factor 2 difference between the values of power in our treatment and Brazier's, given by the different definition of Rayleigh test.

We note that Equation~\ref{eq:xsq} is the probability density function of the noncentral chi squared distribution with two degrees of freedom. It can be shown \cite{grothProbabilityDistributionsRelated1975,brazierConfidenceIntervalsRayleigh1994,vaughanSearchesMillisecondPulsations1994}, that the sum of $n$ powers in Leahy normalization follows, just like the $Z^2_n$ (see below), a non-central chi-squared distribution with $2n$ degrees of freedom:

\begin{equation}\label{eq:ncxsq}
p(\pmeas|\psig) = \mathrm{e}^{-(\pmeas+\psig)/2}{\left(\frac{\pmeas}{\psig}\right)}^{(n-1)/2}\,I_{n-1}(\sqrt{\pmeas\psig})
\end{equation}

This distribution is efficiently implemented in statistical libraries such as \texttt{scipy.stats} and this allows to avoid the somewhat costly and numerically unstable approximations from \cite{grothProbabilityDistributionsRelated1975}.
The corresponding cumulative distribution function is
\begin{equation}\label{eq:ncxsq_cdf}
C(\pmeas|\psig) = 1 - Q_n\left(\sqrt{\psig}, \sqrt{\pmeas}\right)
\end{equation}
where $Q_n$ is the Marcum Q-function.

Given an observed signal power, one can calculate confidence limits on the signal power $P_{sig,\alpha}$ (by assuming \pmeas as an estimate for the mean of the distribution in \eref{eq:ncxsq}).
This can be used to estimate corresponding confidence limits on the  pulsed amplitude, given Equation~\ref{eq:psig}.

\section{Methods not based on the FFT}
\subsection{The Rayleigh test and $Z^2_n$ searches}

The \zsq \citep{buccheriSearchPulsedGammaray1983} is a sensitive statistic for pulsed signals whose pulsed profile can be described by a small number $n$ of sinusoidal harmonics.
Let $t_j\,j=1,\dots\,N$ be the times of detection of $N$ photons.
Given a pulse frequency $f$ and its time derivatives $\dot{f},\ddot{f},...$, we define the \textit{phase} of photon $j$ as
\begin{equation}
    \phi_j = \phi_0 + 2\pi f t_j + \frac{1}{2} 2\pi\dot{f} t_j^2 +\dots
\end{equation}
where $\phi_0$ is a reference phase.
If we detect $N$ photons and their pulse phase is $\phi_i$ ($i=1..N$), we define \zsq as
\begin{equation}\label{eq:zn}
\zsq =\dfrac{2}{N} \sum_{k=1}^n \left[{\left(\sum_{j=1}^N \cos k \phi_j\right)}^2 + {\left(\sum_{j=1}^N \sin k \phi_j\right)}^2\right]
\end{equation}
The case of a single harmonic, the $Z^2_1$, is also known as the Rayleigh test \citep{mardiaStatisticsDirectionalData1975,gibsonTransientEmissionUltrahigh1982}.
Given a certain number of trial pulse periods, one can calculate \zsq and select the periods that give the highest value for the statistic.
For uniformly distributed random events, the \zsq statistic follows a \chizsq distribution, while in the presence of signal it follows the same non-central \chizsq of the sum of $n$ periodograms (Equation \ref{eq:ncxsq}).

The \zsq statistic can also be calculated for binned data \citep{bachettiExtendingStatisticsGeneric2021}.
If $p_j$ is the number of photons detected in bin $j$ and $\phi_j$ their phase, we get
\begin{equation}\label{eq:znbin}
\zsqbin = \dfrac{2}{\sum_i{p_i}} \sum_{k=1}^n \left[{\left(\sum_{i=1}^{\nbin} p_i \cos k \phi_i\right)}^2 + {\left(\sum_{i=1}^{\nbin} p_i \sin k \phi_i\right)}^2\right]
\end{equation}

\subsection{The $H$-test}
What if one is performing a blind search, with no prior information on the pulse shape?
Is there a way to run the \zsq for multiple harmonics, and compare meaningfully the results between the searches?
The answer is yes and it can be done through the $H$ test \citep{dejagerHtestProbabilityDistribution2010}.
It is defined as
\begin{equation}\label{eq:H}
H = \mathrm{max (Z^2_m - 4m + 4)},\,m=1,N_{\rm max}
\end{equation}
where $N_{\rm max}$ is the maximum number of harmonics to investigate.
The test compares the results of \zsq done with an increasing number of harmonics, corrected by the expected degrees of freedom of the statistic.
The $m$ corresponding to the maximum is usually indicated with $M$.

Given a certain value of the statistic $H_0$, the probability distribution of the $H$ statistics is such that, approximately\citep{dejagerHtestProbabilityDistribution2010},
\begin{equation}
    p(H>H_0) = e^{-0.4H}
\end{equation}
The incoherent stacking of multiple $H$ searches, corresponding to the sums of multiple exponentially distributed random variables, follows the Erlang-K distribution.

\begin{figure}
    \centering
    \includegraphics[width=\textwidth]{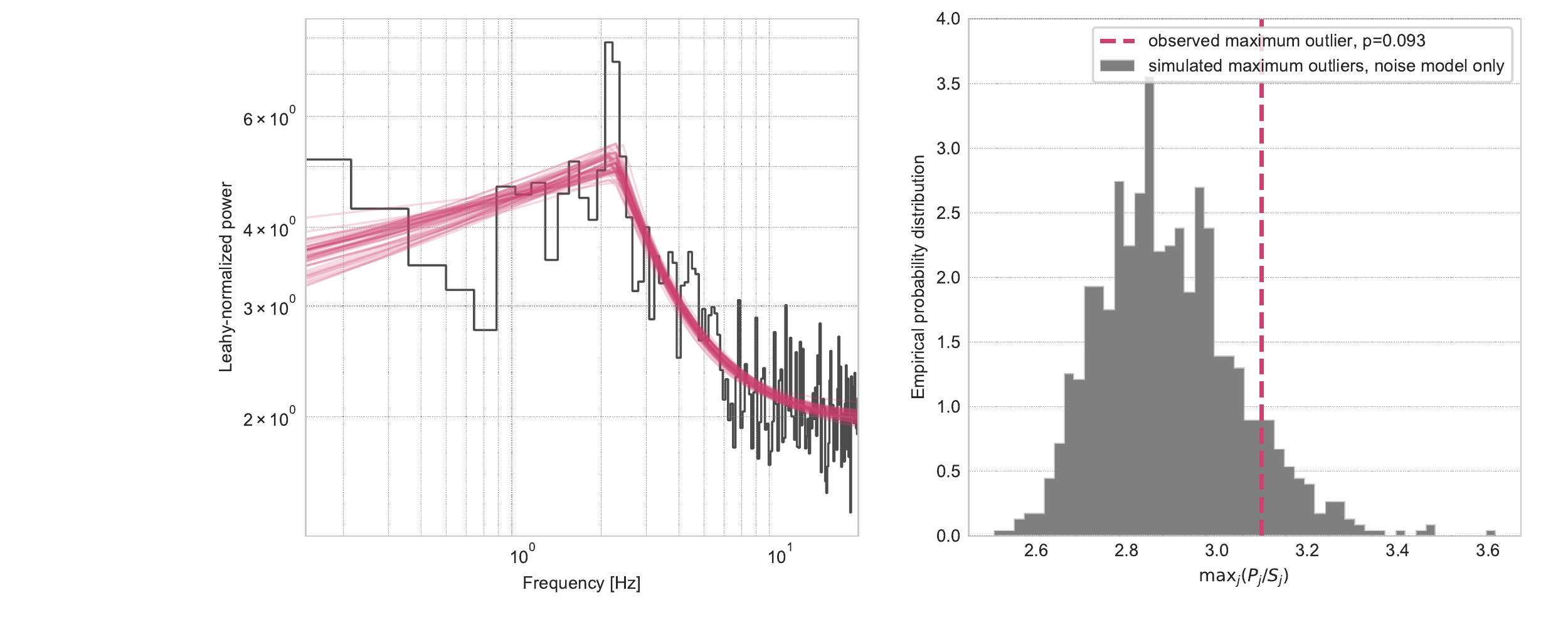}
    \caption{Left: Periodogram of 42 segments of 8 seconds each of the NICER observation of the GRS 1915+105. For demonstration purposes, we limit ourselves to a 300s segment to degrade the signal-to-noise ratio of the strong QPO at 2.4 Hz. The pink lines are draws from a posterior probability density of a broken power law broadband noise model. Right: posterior predictive distribution of the highest outlier statistic defined in Equation \ref{eqn:outlierstat} generated by simulating 1000 periodograms from the broadband noise model posterior distribution. The value of the statistic for the highest outlier in the real data is shown as a pink dashed line, with a posterior predictive p-value of $p=0.093$, which would not be considered significant. The lack of significance is likely due to the fact that the signal is not coherent, and thus spread over multiple neighbouring frequency bins.}
    \label{fig:maxoutlier}
\end{figure}

\subsection{Periodicity Searches in variable light curves: red noise}

One of the fundamental assumptions of the methods we have considered so far is that under the null hypothesis, there is no variability in the light curve. In practice, however, many X-ray sources do show significant variability at a range of frequencies, rendering statistical tests on that null hypothesis misleading. They will find many candidate periodicities across a wide range of frequencies, but the tail probabilities calculated using these tests are effectively meaningless.

\citet{vaughanBayesianTestPeriodic2010} introduced a principled method to search for periodic signals against a red noise process. This process is based on the idea that for a stationary, infinitely long random process, the individual powers in the periodogram still obey a $\chi^2$ distribution, but around the true underlying power spectrum. If the latter is known, one can define a test statistic

\begin{equation}
T_{R,j} = 2 \frac{P_j}{S_j} \;
\label{eqn:outlierstat}
\end{equation}

where $P_j = P(\nu_j)$ is the observed power at frequency $\nu_j$ and $S_j$ is the underlying true power spectrum that generated the data. The test statistic $T_{R,j}$ then has the expected $\chi^2_2$ distribution, and consequently one can use the test against white noise to assess the significance of any candidate $T_{R,j}$ (a good first candidate is, unsurprisingly, $\mathrm{max}(T_{R,j})$).

$S_j$ is not known, but must be inferred from the (usually very noisy) periodogram. If we do not take this uncertainty into account, conclusions from the data will be overconfident. \citet{vaughanBayesianTestPeriodic2010} propose a Bayesian approach that parametrizes $S_j$ (e.g. as a power law or a broken power law), sample the parameters of the model for the broadband noise process, and then generate posterior predictive distributions for $T_j$ in order to assess the significance given the uncertainty in the noise model. This method has been used successfully for example in the study of Active Galactic Nuclei. We show an example on a QPO in GRS 1915+105 observed with NICER \citep{gendreauNeutronStarInterior2012} in Figure \ref{fig:maxoutlier}.

\begin{figure}
    \centering
    \includegraphics[width=\textwidth]{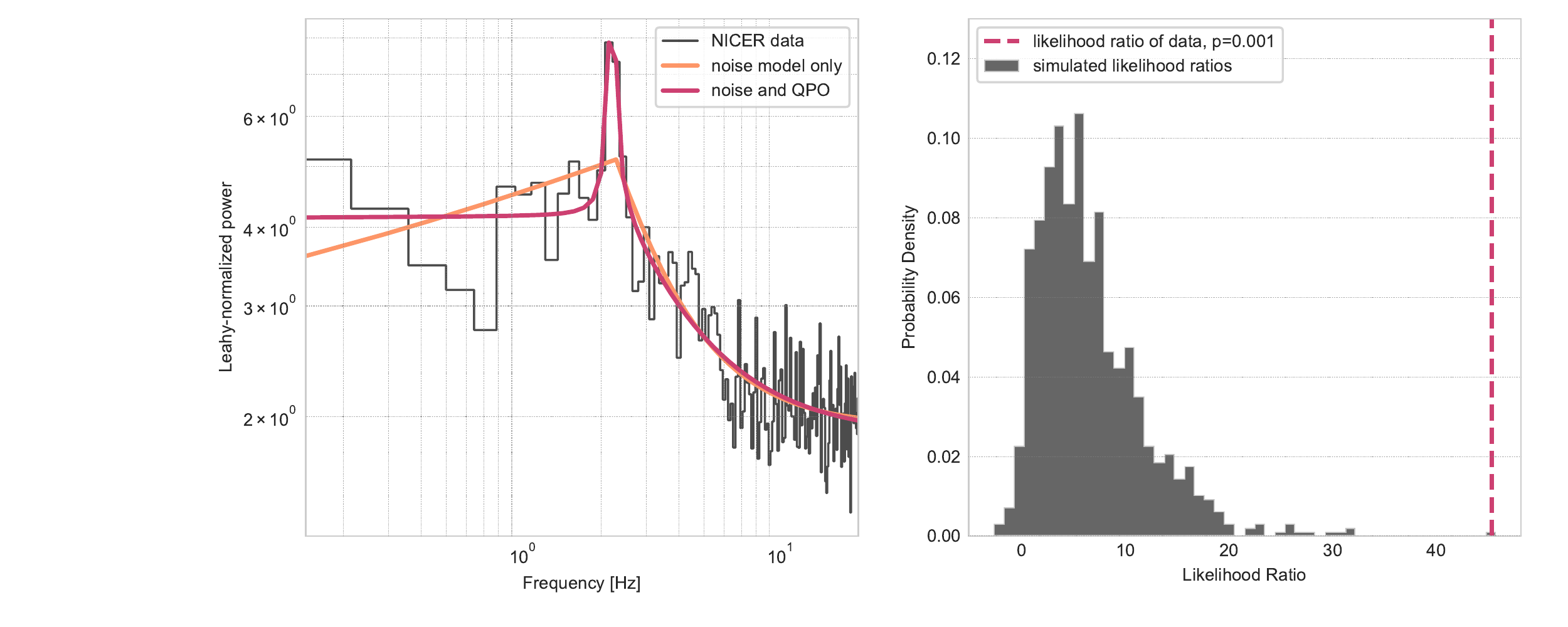}
    \caption{Left: the same periodogram as in Figure \ref{fig:maxoutlier}, with the maximum likelihood models for a broken power-law broadband noise model, and for a combined model of a broken power law and a Lorentzian to model a putative QPO candidate at 2.4 Hz. Right: posterior predictive distribution for the likelihood ratio between the model with a Lorentzian and the one without, for simulated periodograms generated from the broadband noise model only (without QPO). The likelihood ratio of the two models for the observed data is shown as a pink dashed line. Unlike for the outlier statistic, the QPO is clearly significant here (p=0.001), since this test has better sensitivity for quasi-periodic signals.}
    \label{fig:qpo_lrt}
\end{figure}

\subsection{Searching for QPOs with Model Comparison Techniques}

Outlier detection methods generally perform well for systems where the signal to be studied can be expected to be strictly periodic, or at least sufficiently narrow to be contained within a single frequency bin. Because for QPOs, power is often spread across multiple frequencies, outlier detection methods generally have a lower detection sensitivity than for coherent signals.

While these methods can still be successfully applied through the judicious use of rebinning periodograms, it can be difficult to know what binning factors to choose a priori. It can therefore be advantageous to reframe the problem as a model comparison question: can I model my periodogram with only a broadband noise component, or will a combined model that includes a Lorentzian component (or multiple Lorentzians) provide a better model of the observed data?

\citet{vaughanBayesianTestPeriodic2010} provide a method for this, too, which also uses simulations of periodograms generated exclusively from the broadband noise process and assesses significance via posterior predictive distributions. Here, instead of the ratio of periodogram to model, the test statistic is defined via a \textit{likelihood ratio test},

\begin{equation}
    T_{\mathrm{LRT}} = \frac{\hat{\mathcal{L}}(\theta_1)}{\hat{\mathcal{L}}(\theta_0)} \; .
\end{equation}

To construct this test statistic, one fits the broadband noise model $M_0$ (e.g. a power law) to the data via maximum likelihood estimation to obtain the likelihood at maximum, $\hat{\mathcal{L}}(\theta_0)$. One then performs the same optimization for model $M_1$, which generally contains both a broadband noise component and a QPO model to obtain the likelihood at maximum for that compound model, $\hat{\mathcal{L}}(\theta_1)$. The ratio of these quantities only follows a known analytical distribution in certain simple special cases, which are generally not fulfilled in practical problems. Consequently, simulating periodograms from the posterior probability distribution of the broadband noise model, and generating a sample of simulated $T_{\mathrm{LRT}}$ for the null hypothesis where no QPO is present in the data allows the construction of an empirical posterior predictive distribution. Comparing the observed test statistic against this distribution makes it possible to construct a posterior predictive p-value to reject the null hypothesis (no QPO in the data). For an example of this process, see Figure \ref{fig:qpo_lrt}.

\section{Spectral Timing}
\label{sec:spectraltiming}
\begin{figure}
    \centering
    \includegraphics[width=\linewidth]{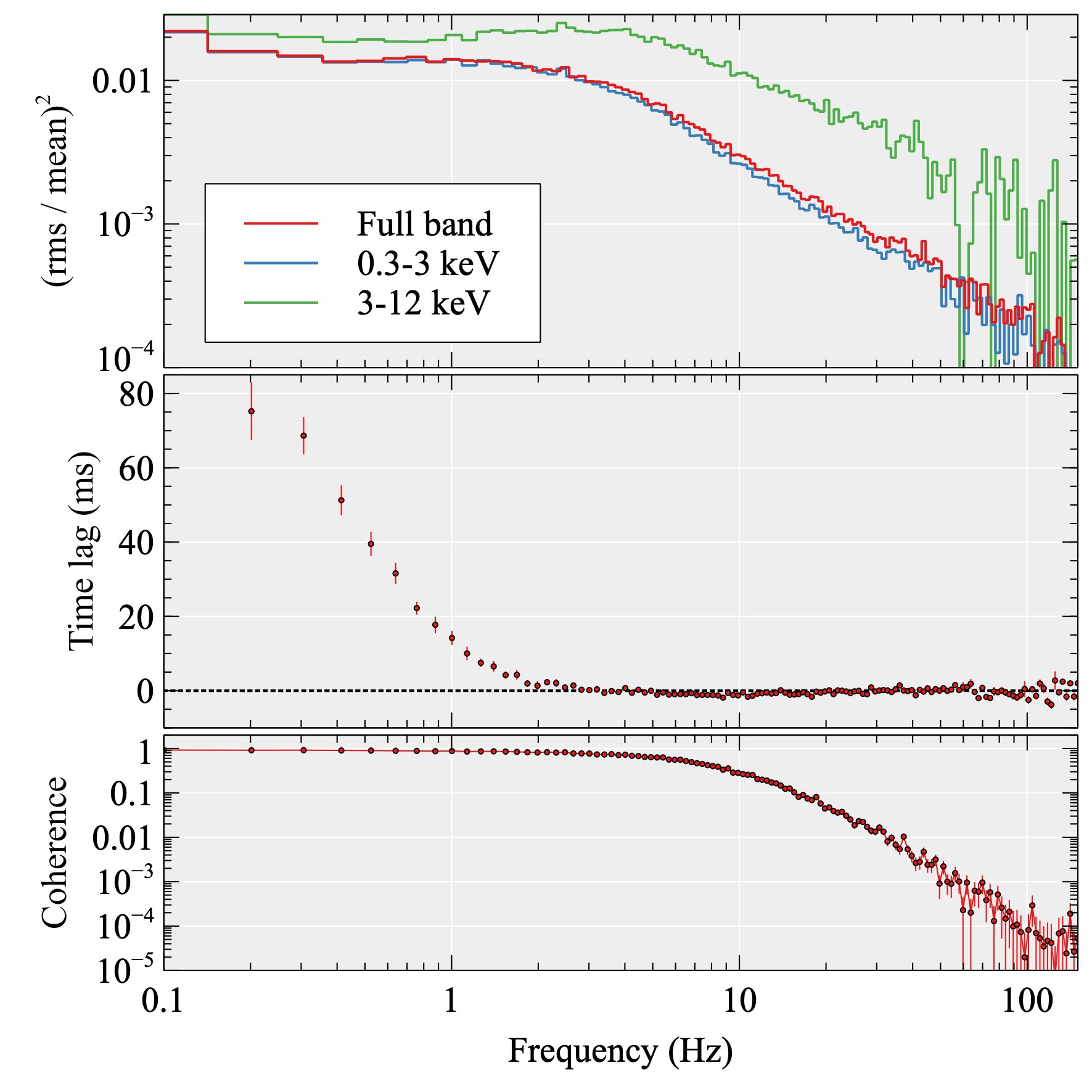}
    \caption{
    (Top) \nicer periodograms of MAXI 1820+070 in its hard state, in different energy bands.
    (Middle) time lags between the 0.3-3 keV and the 3-12 keV band, showing that the low-energy band is leading, i.e. there is a significant hard lag at frequencies 0.1-5 Hz.
    (Bottom) coherence of the cross spectrum between the 0.3-3 keV and the 3-12 keV energy bands
}
    \label{fig:lags}
\end{figure}
The last $\sim$two decades have seen the rise of a class of analysis tools collectively known as \textbf{spectral timing}.
This is a generic term to indicate those techniques that find a connection between the variability in different energy bands.
These are essential, for example, to study the reverberation or propagation of signals in the inner regions of accreting objects \citep{uttleyXrayReverberationAccreting2014b}.
An exhaustive list of all the detailed models developed for different astrophysical cases is beyond the scope of this section, but we will describe here a small set of spectral-timing techniques that form the basic building blocks for these advanced models.

Many of the uncertainties estimated in the cited literature of spectral timing products make a number of assumptions, including but not limited to having intrinsic coherence $\gamma^2=1$ (see below), being in some ``strong signal'' regime, and so on.
\citet{ingramErrorFormulaeEnergydependent2019} derived robust estimates of the uncertainties for spectral timing products, with the only assumption that they were calculated from averaged versions of periodograms and cross spectra using at least $N=40$ single periodograms or adjacent bins.
In many cases, they might be considered overkill, but since they are uniformly better than the others, they do not add excessive complication and can be coded relatively quickly, we encourage to use them instead of alternative formulations.

\subsection{The Cross Spectrum}
Arguably, the most important ingredient to all spectral timing recipes is the cross spectrum.
Let $x_{1, j}, x_{2, j}, j=1,\dots,n$ be two simultaneous and equally sampled time series.
They could be, for example, simultaneous observations of a given source with different detectors, or light curves sampled by the X-rays in different energy bands in the same detector.
Let $a_1(\omega)$ and $a_2(\omega)$ be their Fourier transforms, defined as \eref{eq:ft}.
We define the cross spectrum as
\begin{eqnarray}\label{eq:cross}
    C(\omega) = a_1(\omega)^* a_2(\omega)
\end{eqnarray}
As opposed to the periodogram \eref{eq:periodogram}, this quantity is complex-valued.
Its real part is called the \textit{cospectrum} and its imaginary part is called the \textit{quadrature spectrum}.
If the light curves above only differ in terms of (white) observational noise and there is no delay between them, the cross spectrum can be used as an unbiased periodogram.
This is particularly useful in the cases when we have multiple identical instruments, each affected by deadtime, but independent from one another.
For example, the cospectrum is routinely used by \nustar users as a proxy for the periodogram. \citep{bachettiNoTimeDead2015}.

In order to perform the analysis in the following subsections, we need to make an important assumption.
We need an estimate of the cross-spectrum that is as unbiased as possible, and with Gaussian error bars.
Provided that we are investigating a stationary process, i.e. the statistical properties of the light curves above do not change with time, we can use the cross spectrum $\overline{C(f)}$ obtained from the average of the cross spectra in $N$ segments (similar to Welch's method described above).
Also, we can average together multiple adjacent bins of the averaged cross spectrum, possibly averaging more bins in the (typically higher) frequency range where the signal is weak.
Hereafter, we will assume that all cross spectra and periodograms were obtained as the average of many spectra and bins, dropping the overline symbol.
The choice whether to choose more segments or more averaged bins is largely equivalent.
Defining $N$ the number of original powers summed in each bin (whether from different cross spectra or from different bins), in order to be reasonably sure that error bars can be considered Gaussian, we will need $N\gtrsim40$ \citep{huppenkothenStatisticalPropertiesCospectra2018}.

Until now, we assumed $x_{1, j}$ and $x_{2, j}$ to be independent. In real searches, it might be useful to have the cross spectrum between a light curve in a small energy band $\Delta E_1$ and another inf a larger reference band $\Delta E_2$ which \textit{includes} $E_1$.
This is the case for energy-resolved variability spectra (see below).
In this case, \citet{ingramErrorFormulaeEnergydependent2019} demonstrates that the cross spectrum can be estimated as

\begin{eqnarray}
    C(\omega) = a_1(\omega)^* a_2(\omega) - \mathcal{N}(\omega)
\end{eqnarray}
where the noise term $\mathcal{N}$ turns out to be the white noise level of $P_1$, which is equal to
\begin{eqnarray}
    P_{\rm 1, noise} &= 2\mu &{\rm absolute\,rms\,normalization}\\
    P_{\rm 1, noise} &= \frac{2}{\mu} &{\rm fractional\,rms\,normalization}\\
    P_{\rm 1, noise} &= 2 &{\rm Leahy\,normalization}\\
\end{eqnarray}
where $\mu$ is the mean count rate in the $\Delta E_1$ energy band, in counts per second.

The uncertainty on the real and imaginary parts, the modulus, and the phase of the cross spectrum, in the case where there is \textit{no overlap between the two bands} are \citep{ingramErrorFormulaeEnergydependent2019}:
\begin{eqnarray}
    d\Re(C(f)) =& \sqrt{\dfrac{P_1(f) P_2(f) + [\Re {C(f)}]^2 - [\Im C(f)]^2}{2N}}\\
    d\Im(C(f)) =& \sqrt{\dfrac{P_1(f) P_2(f) + [\Im {C(f)}]^2 - [\Re C(f)]^2}{2N}}\\
    d|C(f)| =& \sqrt{\dfrac{P_1(f) P_2(f)}{N}}\label{eq:dcross}\\
    d|\phi(f)| =& \sqrt{\dfrac{1 - \gamma^2_{\rm raw}(f)}{2\gamma^2_{\rm raw}(f)\,N}}
\end{eqnarray}
where we have introduced the \textit{raw coherence} $\gamma^2_{\rm raw}(f)$, that we will define below.

If the band $E_1$ is included in the reference band $E_2$, the error estimates change as follows:
\begin{eqnarray}
    d\Re(C(f)) =d\Im(C(f)) =d|C(f)| =& \sqrt{\dfrac{P_2}{2N}\left[P_1(f) - \dfrac{|C(f)|^2 - b^2(f)}{P_2 - P_{\rm 2, noise}}\right]}\label{eq:dcrosssub}\\
    d|\phi(f)| =& \sqrt{\dfrac{P_2}{2N}\left[\dfrac{P_1(f)}{|C(f)|^2 - b^2(f)} - \dfrac{1}{P_2 - P_{\rm 2, noise}}\right]}
\end{eqnarray}
where $b^2(f)$ is a \textit{bias term} that we will define below.

\subsection{Coherence}
Coherence measures the amount of linear correlation between two time series.
The literature has multiple definitions of coherence, we will give here the most common ones.
Given two ideal time series $x_{1, j}, x_{2, j}, j=1,\dots,n$ with no noise, the coherence would be \citep{bendatDecompositionWaveForces1986}:
\begin{equation}
    \gamma^2(f) = \frac{{|C(f)|}^2}{P_1(f)P_2(f)}
\end{equation}
However, noise introduces multiple biases in this estimate, which shows up both in the numerator and the denominator.
Estimating the intrinsic coherence from real-life periodograms and cross spectra can be tricky.
In some cases, it is safe to subtract bias terms from the numerator and the factors at the denominator.

The first such correction is known as the \textit{raw coherence}.
Given two noise-affected time series with cross spectrum $C(f)$ and periodograms $P_1(f)$ and $P_2(f)$ respectively, the raw coherence is defined as

\begin{equation}\label{eq:rawcoh}
    \gamma_{\rm raw}^2(f) = \frac{{|C(f)|}^2 - b^2}{P_1(f)P_2(f)}
\end{equation}
where $b^2$ is a bias term
\begin{equation}\label{eq:bias}
    b^2 = \frac{P_1P_2 - \gamma^2[P_1 - P_{\rm 1,noise}][P_2 - P_{\rm 2,noise}]}{N}
\end{equation}
All quantities here can be measured from the data, except the intrinsic coherence $\gamma^2$.
It is usually assumed that $\gamma^2=1$ \citep{vaughanXRayVariabilityCoherence1997}.
\citet{ingramErrorFormulaeEnergydependent2019} proposes an iterative method to estimate it from the data (by substituting back and forth between Equations~\ref{eq:rawcoh} and~\ref{eq:bias}).
Alternatively, it can be estimated by subtracting all noise terms:
\begin{equation}\label{eq:rawcohest}
    \gamma_{\rm est}^2(f) = \frac{{|C(f)|}^2 - b^2}{[P_1 - P_{\rm 1,noise}][P_2 - P_{\rm 2,noise}]}
\end{equation}
However, this can quickly lead to unstable results in energy bands with low count rates, or in frequency bands with low variability.

\subsection{Time lags}
Another important quantity that can be derived by the cross spectrum is called the \textit{phase lag}.
It is the phase angle of the complex-valued cross-spectrum.

Let $x(t)$ and $y(t)$ be two linearly correlated time series.
Let us also define $x$ as the \textit{reference} light curve.
The signal in $y(t)$ also has an additional phase delay with respect to $x(t)$, dependent on frequency $\psi(f)$.
Their Fourier transforms will have the form $X(f) = |X(f)|e^{i\phi(f)}$ and $Y(f)=|Y(f)|e^{i(\phi(f) + \psi(f))}$ respectively.
Therefore, their cross spectrum can be expressed as
\begin{equation}
    C(f) = X^*(f)Y(f) = |Y(f)||X(f)|e^{i\psi(f)}
\end{equation}
In this form, it is clear that the argument of $C(f)$ gives the frequency-dependent phase delay $\psi(f)$ between the two time series. In principle, this quantity can then be used to measure the time delay, or \textit{time lag}, at a given frequency:
\begin{equation}\label{eq:timelag}
    \tau(f) = \frac{\psi(f)}{2 \pi f}
\end{equation}
However, in the presence of noise, this phase is expected to vary randomly, because in this case the Fourier transforms $X(f)$ and $Y(f)$ are random variates themselves.

A single noisy realization of $x(t)$ and $y(t)$ is, therefore, not sufficient to measure the time lag between the two.
Again, it is important to stress that, also for time lags, we need $C(f)$ to be the result of the average of many cross spectra and bins.

From $d|\phi(f)|$ calculated above, it is trivial to estimate the error on the time lag:
\begin{equation}
    d\tau(f) = \dfrac{d\psi(f)}{2\pi f}
\end{equation}

\subsection{Total rms}
An easy measure of the total variability in a given frequency range is given by the total rms.
It is straightforward to measure once we have the averaged periodogram expressed in absolute rms or fractional rms \citep{revnivtsevFrequencyResolvedSpectroscopy1999a}.
\begin{equation}
  \sigma (f) = \sqrt{[P(f) - P_{\rm noise}]\Delta f}
\end{equation}
where $\Delta f$ is the width of the frequency bin where the rms is calculated.

With this definition, the error on the rms spectrum of the reference band would be $\sigma / (2\sqrt{n})$; if calculating the error to fit models to multiple subject bands, however, one would need to account for their correlations. The error formula for each subject band would thus be corrected as  \citep{ingramErrorFormulaeEnergydependent2019}:
\begin{equation}
    d\sigma(f) = \sqrt{\dfrac{[1 - \gamma^4(f)]\sigma^4(f) + \sigma^4_{\rm noise} + 2\sigma(f)\sigma_{\rm noise}}{4N\sigma^2(f)}}
\end{equation}
where $\sigma_{\rm noise}(f)=\sqrt{\Delta f P_{\rm noise}}$

\subsection{Covariance}
A quantity related to the total rms is the \textit{covariance}.
It measures to what extent the variability from two light curves is correlated.
It can be defined either as complex covariance \citep[e.g.][]{mastroserioMultitimescaleXrayReverberation2018}:
\begin{equation}
    {\rm Cov (f)} = \frac{C(f)\,\sqrt{\Delta f}}{\sqrt{[P_2(f) - P_{2,\rm noise}]}}
\end{equation}
or, more commonly, as a real quantity \citep[e.g.][]{wilkinsonAccretionDiscVariability2009}:
\begin{equation}
    {\rm Cov (f)} = \frac{|C(f)|\,\sqrt{\Delta f}}{\sqrt{[P_2(f) - P_{2,\rm noise}]}}
\end{equation}

In the special case where variability at all frequencies and energies is correlated, the covariance spectrum is largely equivalent to the total rms.
However, for non-perfect coherence, it turns out that
\begin{equation}
    {\rm Cov (f)} \approx \gamma(f)\,\mathrm{rms}(f)
\end{equation}

The error bars on the covariance can be calculated by multiplying Equations~\ref{eq:dcross} or~\ref{eq:dcrosssub} (depending on the two energy bands being separate and/or independent) by the factor $\sqrt{\Delta f / (P_2 - P_{\rm 2, noise})}$.

\subsection{Variability-energy spectra}

\begin{figure}
    \centering
    \includegraphics[]{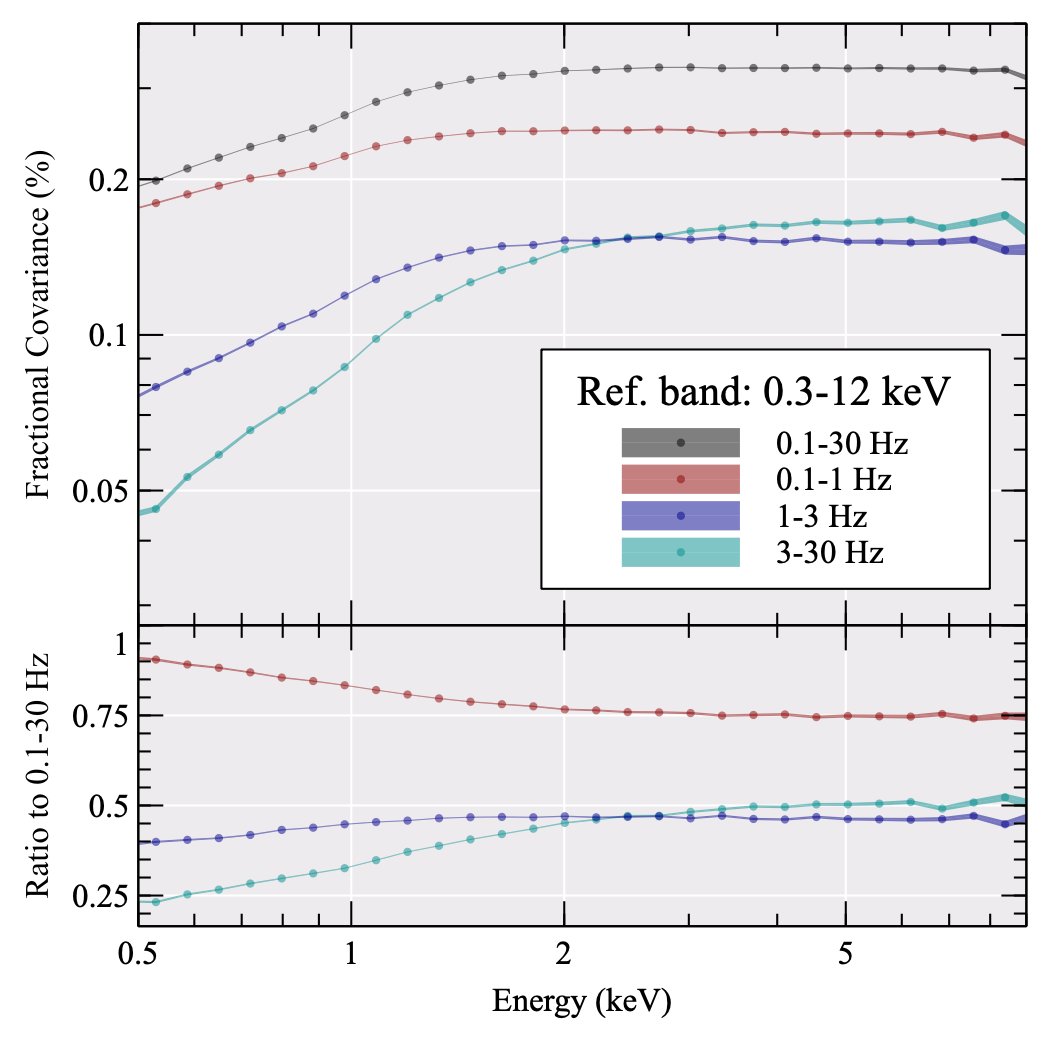}
    \caption{
    (Top) \nicer covariance-energy spectra of MAXI 1820+070 in its hard state, in different \textit{frequency} bands.
    The lower panel shows the ratio of the covariance to the covariance in the 0.1-30 Hz frequency band.
    The reference band was 0.3-12 keV
    }
    \label{fig:covariance}
\end{figure}
\begin{figure}
    \centering
    \includegraphics[]{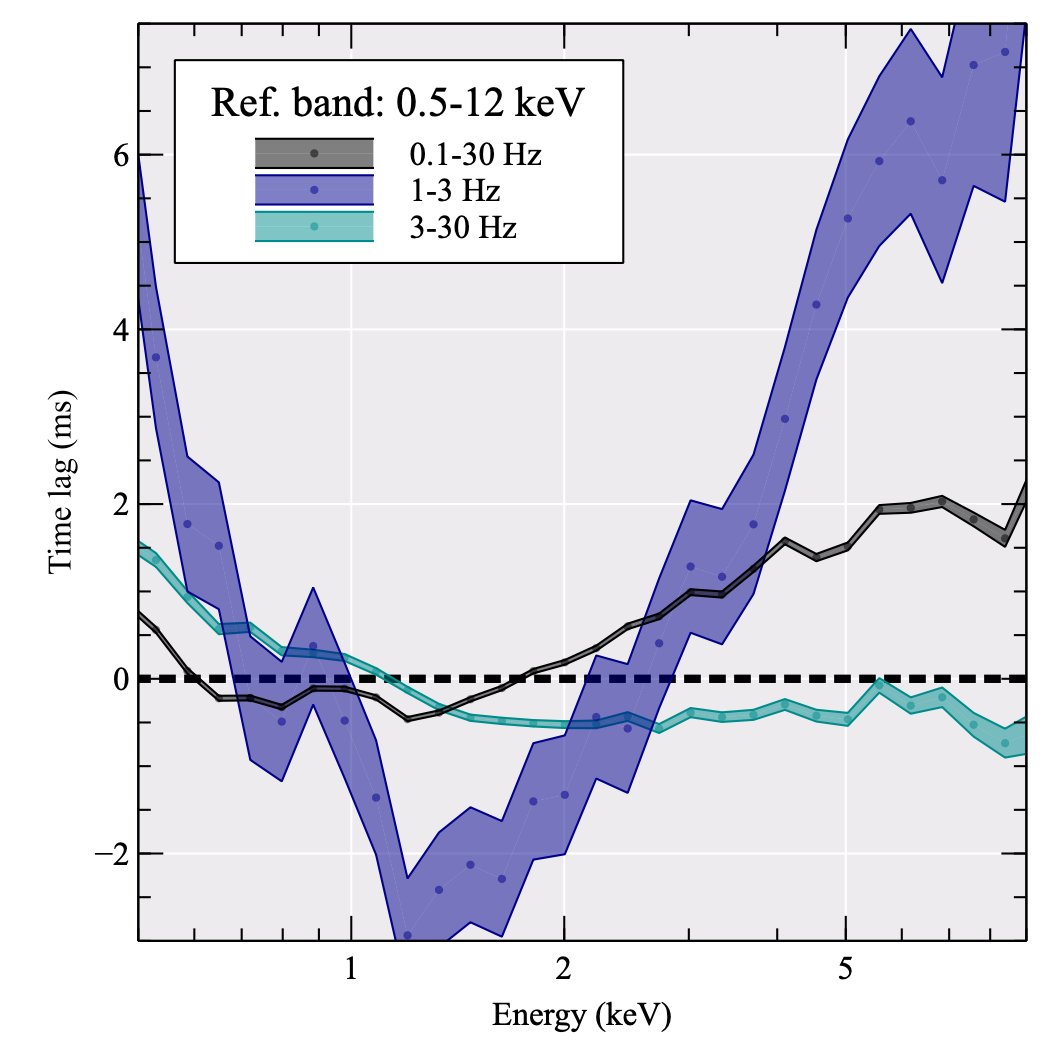}
    \caption{(Top) \nicer lag-energy spectra of MAXI 1820+070 in its hard state, in different \textit{frequency} bands.
    The reference band was 0.3-12 keV}
    \label{fig:lagenergy}
\end{figure}

\begin{figure}
    \centering
    \includegraphics[]{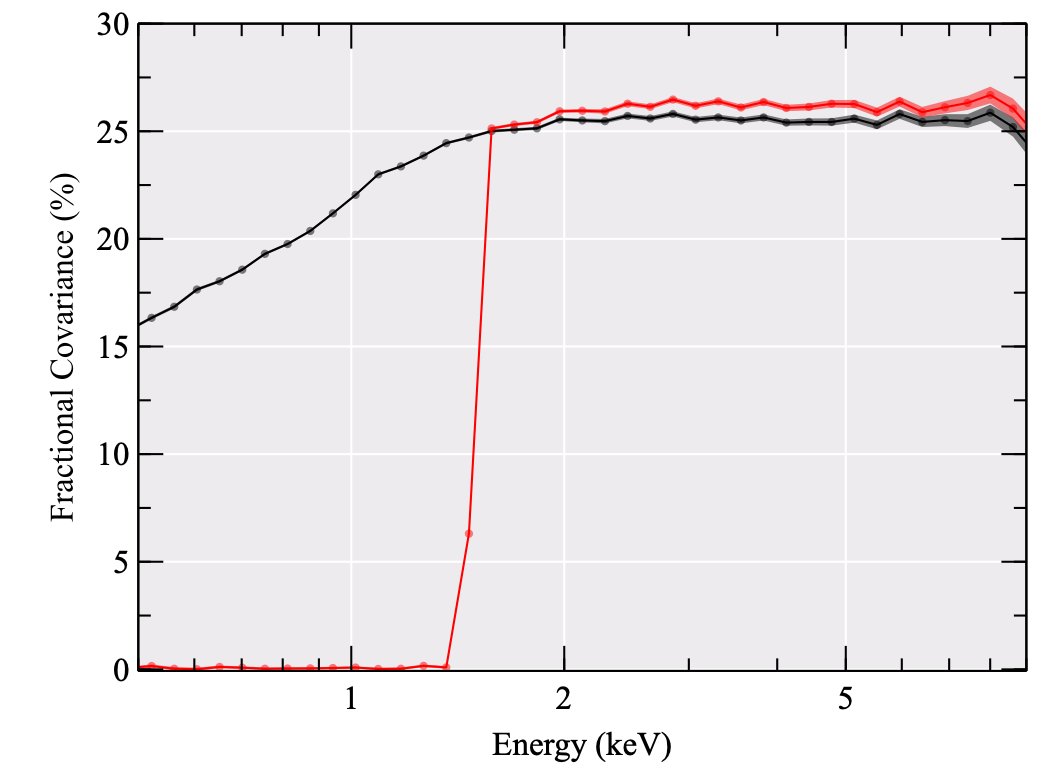}
    \caption{
    \nicer covariance-energy spectrum of MAXI 1820+070, calculated between 0.1 and 5~Hz. The black curve shows the unaltered spectrum. The red curve shows what would happen in the hypothesis that there were no variability below 1.5 keV (we suppressed all variability at energies below 1.5 keV by randomizing the event times).
    The reference band was fixed at 1.5-12 keV}
    \label{fig:cov_suppress}
\end{figure}
Once one has a working infrastructure to calculate basic building blocks such as cross spectra, periodograms, and time lags, it is possible to use them to study the energy-resolved variability, or the spectrum of single variability components.

Most of these advanced spectral timing products work along a similar procedure (see \citealt{uttleyXrayReverberationAccreting2014b}):

\begin{enumerate}
    \item select a number of \textit{channels of interest} (CI), energy ranges at which a given variability measurement will be made;
    \item select a \textit{reference energy band}, which needs to be a band where variability can be measured significantly; it is possible to use the full band as a reference band, see caveats below;
    \item extract event lists from each CI and the reference band, and produce light curves
    \item select the \textit{frequency band} the variability needs to be evaluated at
    \item for each energy band, calculate averaged cross spectrum and periodograms as needed to measure the wanted quantity (time lag, rms, covariance, etc.). The reference band is the one complex conjugated in \eref{eq:cross}, if the sign convention of the Fourier transform is the same as \eref{eq:ft}, otherwise the two bands need to be transposed;
    \item average over the wanted frequency band(s)
    \item calculate time lags, covariance, and any other derived quantity
    \item evaluate the uncertainties
\end{enumerate}
We would like to stress here that the order of operations above is very important.
It is always more appropriate to average cross spectra and then calculate, e.g., lags, rather than averaging the calculated lags.
Also, when the frequency bands are wide, the choice of what frequency to use in \eref{eq:timelag} must be done carefully: different authors might choose different solutions based on their use case (e.g. the simple mean, a geometrical mean, a mean weighted by the cross spectral amplitudes), but whatever the choice, it should always be clearly specified.
If possible, any modeling of the lags should be done using phase lags instead of time lags.

The overlap between the CI and the reference band can in principle create a bias in these measurements.
Many authors (e.g. \citealt{uttleyXrayReverberationAccreting2014b}) recommend to clean up the reference band from the photons used in each CI when evaluating spectral-timing quantities.
This, however, means that the reference band will be slightly different for each CI.
As we anticipated when talking about the cross spectrum, \citet{ingramErrorFormulaeEnergydependent2019} takes an alternative approach: he demonstrates that the cross spectrum evaluated between a reference band and an overlapping CI is only biased by an amount corresponding to the white noise term of the CI periodogram, and so this noise term just needs to be subtracted from the cross spectrum.

In Figures~\ref{fig:lagenergy} and \ref{fig:covariance} we show two examples of variability-energy spectra using the first NICER observation from the hard state of the black hole binary MAXI 1820+070 during its 2018 outburst.
Figure~\ref{fig:lagenergy} shows the time lags at different Fourier frequencies for the black hole binary MAXI 1820+070, largely reproducing the results from \citep{karaCoronaContractsBlackhole2019,demarcoInnerFlowGeometry2021,wangDiskCoronaJet2021}.
Figure~\ref{fig:covariance} shows the covariance spectrum from the same observation, in the same energy bands, and the ratio between the covariance spectra at different frequencies, showing the change of slope.
Figure~\ref{fig:cov_suppress} shows how the covariance spectrum would change if, for some reason, variability below 1.5 keV were suppressed.

\section{Common Pitfalls}
\label{sec:pitfalls}
\subsection{Detector effects: Dead Time and friends}

\begin{figure}
    \centering
    \includegraphics[width=\textwidth]{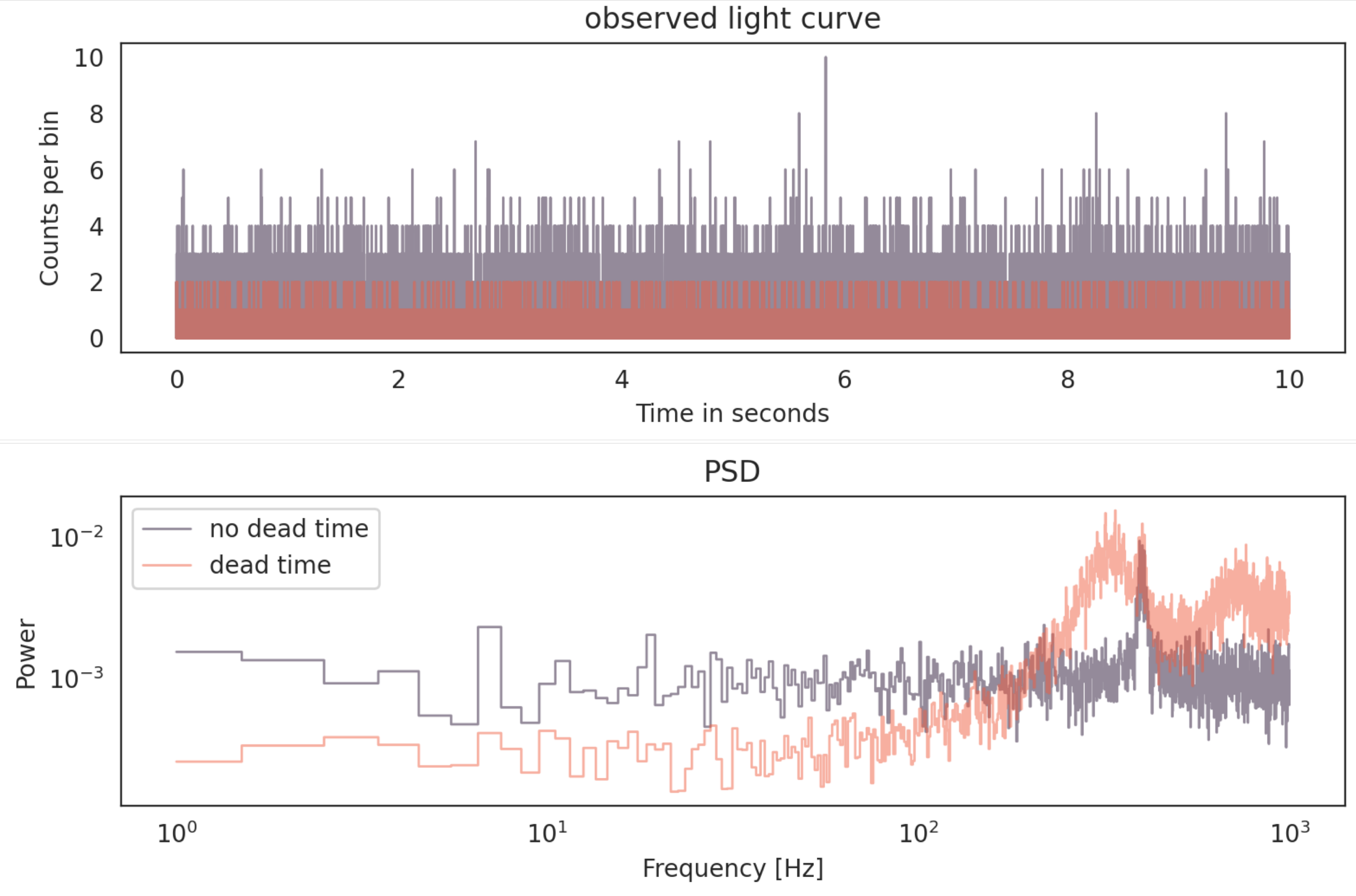}
    \caption{Simulated example of NuSTAR-like dead time. Top: light curve of a process without dead time (purple) and with dead time (orange). Because dead time effectively ignores a subset of photons during the detection process, the light curve with dead time has a lower mean count rate than the one without. Bottom: the periodograms corresponding to the two light curves in the top panel make it apparent that dead time also alters the frequency response of the periodogram.}
    \label{fig:my_label}
\end{figure}

X-ray and gamma-ray detectors often suffer from an effect called \textit{dead time}.
This is the time \deadt that the detector chain stops in order after recording an event, to check that it corresponds to a valid photon (and not, for example, a cosmic ray) and to measure its energy, position, and any other physical properties.
There are two main kinds of dead time, depending on what happens when another photon hits the detector during the previous photon's dead time.
If the detector chain initiates another dead time cycle from the new event, at high count rates this might potentially lead to completely ``freezing'' the detector. This is called \textit{paralyzable} dead time. Detectors affected by paralyzable dead time, like the upcoming Athena/X-IFU \citep{barretATHENAXrayIntegral2018}, cannot be used for observations of very bright sources unless mitigation strategies are implemented (such as the defocusing over multiple independent detectors).
When all photons hitting the detector during dead time are simply ignored, dead time is called \textit{non-paralyzable}.
This is the case for most astrophysical missions, from \xte \citep{jahodaCalibrationRossiXRay2006} to \nustar \citep{harrisonNuclearSpectroscopicTelescope2013a}.
Dead time creates correlations between events, that are therefore not independent anymore.
This produces a number of effects on the temporal properties of the signal, some immediately visible, some subtle:
\begin{itemize}
    \item the white noise level of the periodogram is not constant, but becomes ``wavy'' with saddle points at frequencies $n/(2\deadt)$
    \item the variance of the periodogram is itself modulated, with the same wavy behavior of the white noise level.
    \item cross products are also affected, and lag measurements are dominated by features at multiples of the dead time.
\end{itemize}
In some missions like \xte, dead time is so short that its effect can be modeled as a constant offset of the white noise level.
Others, like \nustar, need mitigation strategies in order to allow the characterization of variability above $\sim10$\,Hz.
In principle, the effect of dead time on the periodogram can be modeled, provided that dead time is constant \citep{vikhlininQuasiperiodicOscillationsShot1994,zhangDeadTimeModificationsFast1995a}.
The constant-dead time modeling might, in some cases, be adequate.
However, this is rarely the case.
Dead time depends on the time that the detector takes to process an event, and this in turn might depend on the amount of pixels the event was recorded in (the so-called event \textit{grade}), and other subtleties of the spacecraft detector chain.
For example, \ixpe's Gas Pixel Detectors (GPD) measure the energy and polarization of incident photons through the track that a photoelectron leaves in the pixels \citep{weisskopfImagingXrayPolarimetry2016}.
The processing depends on the number of pixels touched by the photoelectron, and in particular this is dependent on the energy of the photon.
Hence, the dead time is proportional to the photon energy, and this is likely to affect spectral timing products.

When the mission carries multiple, independent detectors, the effects of dead time can be mitigated in a relatively simple manner.
Time lags between different energy bands, for example, can be calculated using photons from one detector for the reference band and from the other detector for the band of interest.
Provided there is no significant contribution from events hitting multiple detectors at once (e.g. the Very Large Events in \xte), this method yields lag measurements that should be mostly unaffected by dead time.
The cospectrum between the signal in the two detectors, instead, can in principle be used as a white-noise-subtracted proxy of the periodogram \citep{bachettiNoTimeDead2015}.
However, the variance of white noise in the dead time-affected cospectrum is not constant, and this only shifts the issue of dead time from the detection over white noise to the interpretation of the significance of a given feature in the periodogram with respect to the local variance.
\citet{bachettiNoTimeDead2018} devised a powerful method to correct the white noise level of averaged periodograms and cospectra from dead time, using the difference between the complex Fourier amplitudes of the signals from the two detectors (the Frequency Amplitude Difference, or FAD, method).

\citet{huppenkothenAccurateXrayTiming2021} use a Simulation-Based Inference (SBI) approach to characterize the variability in dead time-affected periodograms, also when there is only one detector or the FAD cannot be applied.
This method does not try to mitigate dead time or model it directly.
Instead, it tries to to recover a Bayesian posterior probability distribution of the parameters of variability components from simulations, given a well-designed metric to measure the distance between the observed and simulated data.
The principle is simple: instead of correcting the periodogram for dead time, one simulates dead time-affected data with the variability components they want to model, and tries to infer the posterior distribution of the parameters by looking at the distance between the simulated and the real periodograms.
The simulation can be done, in principle, using the \citet{timmerGeneratingPowerLaw1995b} method, transforming the light curve into photons, applying dead time, and calculating the periodogram for each step.
In practice, doing this by brute force would be prohibitive.
\citet{huppenkothenAccurateXrayTiming2021} investigate various machine learning techniques based on neural networks to train a simulator that is able to produce synthetic periodograms in order to perform the inference.

\subsection{Non-stationarity}

The statistical validity of the Fourier methods outlined in this Chapter generally rest on fairly strong assumptions: an infinitely long time series of evenly spaced measurements without gaps, generated by a stationary process. These can obviously never be fully met, given the finite observing time available on our telescopes. However, for practical purposes, what is often more important than strictly meeting all of these requirements is to make sure they are fulfilled \textit{with respect to the frequencies of interest}. For example, a black hole X-ray binary in outburst will show non-stationary changes in its light curve over the course of days to weeks. But frequencies of physically relevant QPOs might be in the range of a few Hz to a few hundred Hz, often easily recognizable within a single observation, where the assumption of stationarity may be adequate.

Similarly, the rapid, non-stationary variability of magnetar bursts may not be important when searching for QPOs at frequencies as high as $1600\mathrm{Hz}$ as have been observed in the magnetar giant flares \citep{israel2005,strohmayer2006,watts2007}. At these frequencies, one may effectively compare to white noise only, and the variability at low frequencies bears only negligible effects on the detection probabilities. Care needs to be taken when pushing Fourier methods into the limits to the edge of their capabilities, or into regimes where some of the underlying assumptions might be broken in ways that significantly affect the statistical properties.

For the purposes of this chapter, we will consider two specific cases: (1) a (stationary) quasi-periodic signal on top of a transient light curve, and (2) a non-stationary QPO signal in a stationary or transient light curve. The former case defines the QPO search problem in fast transients such as GRBs \citep{cenko2010} and magnetar bursts \citep{huppenkothen2013,huppenkothen2014}. The latter might occur when a QPO might not be present over the entire light curve (or segment of interest).

Detecting QPOs in transients constitutes a particularly complex case for the detection of periodic and quasi-periodic processes, because their inherent non-stationary nature break some of the fundamental assumptions of the common Fourier-based statistical methods introduced in this Chapter. If the expected frequencies of the QPO are high compared to the overall duration of the burst, and if these signals are well-separated in frequency space from other variability, then standard Fourier methods may be a good match to the problem. In this case, the assumptions of stationarity (and often, the even stronger assumption of white noise) can be a reasonable model for the observations.

\citet{huppenkothen2013} extended the work in \citet{vaughanBayesianTestPeriodic2010} to work with short magnetar bursts. Similarly, \citet{inglis2015} build a framework based on the same work for automated detection of QPOs in solar flares. Both assume that red noise is a reasonable assumption for the variability observed in magnetar bursts.

For short magnetar bursts with durations of the order of $\sim0.1 - 1$s, validation experiments show that detection of periodic signals using the assumption of red noise becomes very difficult below $\sim 40\mathrm{Hz}$ or so. At low frequencies, the window function imposed by the finite nature of the burst imposes structure onto the light curve that is statistically not well modelled by a red noise process. If the underlying shape of the light curve is well-modelled by a physically motivated deterministic process, it should be possible to design a detection procedure that incorporates that knowledge. For many transients like magnetar bursts and GRBs, however, this is often not the case.
Recently, \citet{huebner2022} have shown that when searching for a non-stationary QPOs against a variable background light curve, standard methods strongly overestimate the statistical significance of a QPO candidate, and that the significance can depend on particular choices made in the data preparation. Thus care must be taken when either the QPO or the light curve as a whole are likely to be non-stationary with on timescales relevant to the expected QPO frequency.

\subsection{Unevenly Sampled Data: The Lomb-Scargle Periodogram}

One of the strongest assumptions about the data that traditional Fourier methods make is that of even sampling. In X-ray astronomy, this is often not a bad assumption, especially for instruments that record individual photon arrival times: because the sampling of the time series can be more or less freely chosen by the researcher, it can be made to be strictly even (up to machine precision). However, this will only hold for studies where the frequency range of relevance is covered by a single, uninterrupted observation. This can severely limit studies in practice: even within relatively long X-ray observations, gaps will occur due to the spacecraft's motion (and its occasional occlusion by Earth). Similarly, studies of long-term timing behaviour utilize time series consisting of individual observations taken over months or years, which are subject to observing and scheduling constraints. The latter impose a highly irregular sampling function onto the data. If this sampling function were entirely random, its convolution with the underlying signal would effectively impose another source of stochastic noise onto the periodogram. If the sampling function is somewhat regular, then that regularity imposes timescales on the data that will convolve with the source signal to produce aliases that are difficult to predict and equally difficult to deconvolve. While observations from space are often subject to somewhat less stringent observing constraints than from the ground (due to the lack of day/night rhythm and seasonality), observing gaps often do occur on the fairly regular orbital period of the satellite or due to other, spacecraft-specific constraints.

The solution to this problem again depends on the frequency range of interest and the precise structure of the data. X-ray telescopes often produce evenly sampled light curves with gaps during intervals where the Earth is between spacecraft and source. For frequencies that are short compared to the typical observing window between occultations, splitting the light curve into segments and generating Fourier periodograms for each segment might be sufficient. For signals on longer time scales or uneven sampling patterns, other methods exist for timing studies that can better account for the structure of the data, including Lomb-Scargle periodograms (\citealt{lomb1976,scargle1982}; for a comprehensive review, see also \citealt{vanderplas2018}), Gaussian Processes \citep{wilkins2019}, continuous autoregressive moving average processes (CARMA; \citealt{kelly2014}), and wavelets \citep{foster1996}. There has been limited work on spectral timing with unevenly sampled data, for example for time lags \citep{zoghbiCALCULATINGTIMELAGS2013}, but this is currently an active field of research.

One key issue of generating a Schuster periodogram using data with non-uniform sampling is that our guarantees for the statistical distributions we can expect for typical noise processes (especially white noise and red noise) are only correct in the limit of even sampling. The presence of a non-uniform sampling function leads to periodograms for which the statistical distribution is not generally analytically tractable. This, in turn, complicates both periodicity searches and modeling of the periodogram. As a solution to this issue, \citet{scargle1982} suggested a generalized version of the periodogram appropriate for unevenly sampled data:

\begin{equation}
\begin{split}
    P_\mathrm{LS}(\nu) = &\frac{1}{2}\left[   \left( \sum_{i} x_i \cos{2 \pi \nu (t_i - \tau)}\right)^2 / \sum_i\cos^2{2\pi \nu (t_i - \tau)} + \right. \\
    & \phantom{BLA}\left.\left( \sum_{i} x_i \sin{2 \pi \nu (t_i - \tau)}\right)^2 / \sum_i\sin^2 2\pi \nu (t_i - \tau) \right]
\end{split}
\end{equation}

\noindent with
\[
\tau = \frac{1}{4 \pi \nu} \tan^{-1}\left( \frac{\sum_i \sin(4\pi \nu t_i)}{\sum_i \cos(4\pi \nu t_i)}  \right) \; .
\]

\noindent Interestingly, this formulation of the periodogram corresponds to a maximum likelihood solution for an ensemble of sinusoidal basis functions at frequencies $\{\nu_k\}_{k=1}^{N/2}$ fit to the observations using a Gaussian likelihood function, first explored by \citet{lomb1976}. That is, if one fit a sinusoidal function to the time series with frequency $\nu_k$ and free parameters $A_k$ and $\phi_k$, and then obtained the maximum likelihood solution, the resulting set of maximum likelihood values as a function of frequency $\nu$ correspond to the \textit{Lomb-Scargle periodogram} (LSP) as defined above. As with the classical periodogram, the powers in the LSP follow a $\chi^2$ distribution with 2 degrees of freedom. Note, however, that the choice of frequency grid is less well defined as in the regularly sampled case, because there is no straightforward analog to the Nyquist frequency determining the maximum frequency that should be considered. We refer the reader to \citet{vanderplas2018} for  more information about the LSP and idiosyncrasies.

\section{Conclusions}
In this chapter, we have provided an overview of several Fourier-based techniques utilized in high-energy astronomy to study variability. These techniques have demonstrated their ability to provide profound insights into the physics of observed objects, particularly when considering the complete information, including the relationship between emission patterns at different energies. Although these techniques are increasingly popular due to the availability of open-source software packages, it is important to acknowledge that they rely on certain assumptions (such as stationarity and uniform instrumental response) that may not always hold true in real observations. We have made an effort to explicitly address these caveats and have also presented strategies to mitigate instrumental systematics, such as the impact of dead time or the presence of missing or non-uniformly sampled data.

\section{Acknowledgements}
The authors wish to thank the referee Phil Uttley for comments and suggestions that really improved the quality of this Chapter, the Editors of the Handbook Cosimo Bambi and Andrea Santangelo, and the Editors of this Section Tomaso Belloni and Dipankar Bhattacharya for their coordination.
They also thank Simon Vaughan for very insightful comments on the manuscript, and Guglielmo Mastroserio, Barbara De Marco, Erin Kara, and Jingyi Wang for useful discussions on spectral timing.

MB wishes to acknowledge funding from PRIN TEC INAF 2019-``Spectempolar!''.
DH is supported by the Women In Science Excel (WISE) programme of the Netherlands Organisation for Scientific Research (NWO).
The analysis and the plots in this paper were done using Open Source software, including but not limited to \texttt{astropy} \citep{astropycollaborationAstropyProjectBuilding2018a}, \texttt{Stingray} \citep{huppenkothenStingraySpectraltimingSoftware2016,huppenkothenStingrayModernPython2019}, \texttt{HENDRICS} \citep{bachettiHENDRICSHighENergy2018}, \texttt{scipy} \citep{2020SciPy-NMeth}, \texttt{numpy} \citep{harris2020array}, \texttt{numba} \citep{numba2015}, \texttt{Veusz}\footnote{https://veusz.github.io/}, \texttt{Matplotlib} \citep{Hunter:2007}, \texttt{corner} \citep{corner}, \texttt{emcee} \citep{foreman-mackeyEmceeMCMCHammer2013}.

The examples make use of data and software provided by the High Energy Astrophysics Science Archive Research Center (HEASARC), which is a service of the Astrophysics Science Division at NASA/GSFC.

\section{Bibliography}
\bibliographystyle{spbasic}
\bibliography{fourier}

\end{document}